\documentclass[12pt]{article}
\pdfoutput=1
\usepackage{jheppub}
\usepackage{amsmath}
\usepackage{bm}
\usepackage{slashed}
\usepackage{graphicx}

\newlength{\dummysp}
\newcommand{\beq}{\begin{eqnarray}}
\newcommand{\eeq}{\end{eqnarray}}

\newcommand{\gappeq}{\mathrel{\rlap {\raise.5ex\hbox{$>$}}
{\lower.5ex\hbox{$\sim$}}}}
\newcommand{\lappeq}{\mathrel{\rlap{\raise.5ex\hbox{$<$}}
{\lower.5ex\hbox{$\sim$}}}}

\newcommand{\ben}{\begin{enumerate}}
\newcommand{\een}{\end{enumerate}}

\newcommand{\bit}{\begin{itemize}}
\newcommand{\eit}{\end{itemize}}

\def\[{\left [}
\def\]{\right ]}
\def\({\left (}
\def\){\right )}

\title{$S$-duality, boundary states, and higher-form symmetries on ALE spaces}
   
 \author{Mohamed M. Anber}
  
\affiliation{Centre for Particle Theory, Department of Mathematical Sciences, Durham University, South Road, Durham DH1 3LE, UK}

\emailAdd{mohamed.anber@durham.ac.uk}  

\abstract{

{\flushleft{W}}e study Abelian $S$-duality of Maxwell theory on $A$-type asymptotically locally Euclidean (ALE) spaces. Unlike on closed four-manifolds, the Maxwell path integral on an ALE space is not naturally a scalar partition function. Rather, it decomposes into theta-function blocks labeled by flat $U(1)$ holonomy sectors on the asymptotic lens-space boundary. We interpret these blocks as components of the Hilbert-space boundary state prepared by the ALE path integral. With this interpretation, the apparent failure of ordinary modularity is replaced by vector-valued modular covariance under the action of the modular group.
We test this picture explicitly for Eguchi--Hanson space by gluing it to its orientation reversal. The resulting closed four-manifold is diffeomorphic to $S^2\times S^2$, and the natural pairing of the two ALE boundary states reproduces the standard Maxwell partition function on $S^2\times S^2$. We then refine the construction by turning on electric and magnetic $1$-form symmetry backgrounds. In their presence, the ALE theta blocks are not ordinary functions, but sections of a line bundle over the Cartan torus associated with the $A_{N-1}$ root lattice, reflecting the mixed electric-magnetic $1$-form anomaly. We also discuss gauging discrete $\mathbb Z_k$ subgroups of the $1$-form symmetries and show that the vector-valued boundary-state structure remains the natural covariant framework after gauging. In this sense, ALE spaces behave as chiral building blocks for four-dimensional Maxwell theory: individual ALE blocks carry sector-resolved boundary data, while gluing pairs these sectors to produce an ordinary closed-manifold partition function, much like the pairing of left- and right-moving conformal blocks in two-dimensional CFT.
}

\begin{document}

\maketitle

\flushbottom

\section{Introduction}

It is well known that Maxwell theory enjoys electric-magnetic duality, relating an electric description with coupling $g$ to a magnetic description with coupling $g^{-1}$. More generally, this strong-weak coupling duality, usually referred to as $ S$-duality, has appeared in a wide range of settings, from lattice gauge theories \cite{Cardy:1981qy,Cardy:1981fd,Shapere:1988zv,Anosova:2022yqx} and Abelian gauge theory on curved manifolds \cite{Witten:1995gf,Verlinde:1995mz,Metlitski:2015yqa} to the celebrated $S$-duality of ${\cal N}=4$ topologically-twisted super Yang--Mills theory on $4$-manifolds \cite{Vafa:1994tf}. In its most refined form, the duality is not merely a statement about the equations of motion, but about the full quantum theory: the partition function, the spectrum of line operators \cite{Aharony:2013hda}, and the response to background fields transform in a controlled way under the action of the modular group $SL(2,\mathbb Z)$ on the complexified coupling
$
\tau=\frac{\theta}{2\pi}+\frac{4\pi i}{g^2}.
$
Recently, this duality has been scrutinized in the light of generalized symmetries, see, e.g., \cite{Choi:2022rfe, Hayashi:2022fkw,Kaidi:2022uux,Hofman:2024oze,Shao:2025qvf}.

On a closed $4$-manifold, this structure is well understood. The Maxwell path integral is a partition function that transforms under the modular group as a modular form with weights determined by the topology of the underlying manifold \cite{Witten:1995gf,Verlinde:1995mz}. However, the situation becomes more subtle, and in some ways more informative, when the $4$-manifold is noncompact and has an asymptotic boundary. In that case, the path integral should not be regarded simply as a number. Rather, by the usual cutting-and-gluing principles of quantum field theory, it naturally prepares a state in the Hilbert space associated with the boundary $3$-manifold \cite{Atiyah:1989vu}.

Among the structurally-rich noncompact spaces are the asymptotically locally Euclidean (ALE) spaces \cite{Eguchi:1980jx,Kronheimer:1989zs,PeterBKronheimer:1990zmj}. These $4$-manifolds approach a flat-space quotient at infinity. More precisely, outside a compact region, an ALE space is asymptotic to $\mathbb R^4/\Gamma$, where $\Gamma$ is a finite subgroup of $SU(2)$. The simplest examples are the $A_{N-1}$ ALE spaces, which resolve the orbifold singularities of $\mathbb C^2/\mathbb Z_N$. These spaces are hyperkähler manifolds that contain a collection of compact exceptional $2$-cycles whose intersection form is the Cartan matrix of the corresponding ADE Lie algebra.  Asymptotically, they reach a lens-space boundary. This makes them especially useful for studying gauge theory on manifolds with boundary: fluxes can be supported on the compact $2$-cycles, while the lens-space boundary allows for torsion data.

ALE spaces appear in mathematical physics, gauge theory, and string theory. For example, they play an important role in supersymmetric compactifications, geometric engineering, and the study of BPS states, because their hyperkähler structures preserve part of the supersymmetry \cite{Katz:1997eq,Johnson:1996py}. In gauge theory, ALE spaces are central to the study of instanton moduli spaces, notably in Nakajima's construction of affine ADE representations \cite{Nakajima:1994nid}. In quantum gravity and Euclidean field theory, they provide explicit noncompact gravitational instanton backgrounds with nontrivial topology and a lens-space boundary. Recently, the author showed that these solutions can also play a role in gauging the Standard Model $\mathbb Z_6^{[1]}$ $1$-form symmetry \cite{Anber:2025gvb}, and can serve as probes to detect 't Hooft anomalies not seen on conventional closed manifolds \cite{Anber:2025dxy}. Thus, ALE spaces serve as controlled local laboratories for singularity resolution, flux sectors, instantons, and the interplay between geometry and quantum field theory.

The $S$-duality properties of free Maxwell theory on  ALE spaces were checked in \cite{Bianchi:1996zj}. An important observation of that work is that the object obtained by evaluating the Maxwell path integral on an ALE space does not transform as a modular form, in contrast with the familiar behavior on closed $4$-manifolds. This apparent failure is closely tied to the fact that the lattice of homology cycles of ALE space is not self-dual. At first sight, this might suggest that $S$-duality is violated on ALE backgrounds. However, because ALE spaces play a central role in many field-theory and string-theory applications, it is important to understand more precisely what this statement means. In particular, one should ask whether the apparent failure of modularity signals a genuine breakdown of duality, or whether it instead reflects the fact that the ALE path integral is not a closed-manifold partition function, but a more refined object associated with a space with boundary.

The purpose of this work is to revisit this question, especially in light of generalized global symmetries: higher-form symmetries and their background fields \cite{Gaiotto:2014kfa}. We focus throughout on ALE spaces of $A$-type, for which the compact homology lattice is naturally identified with the root lattice of the $A_{N-1}$ Lie algebra. 

An important ingredient in the construction is the existence of normalizable self-dual Maxwell field strengths on ALE spaces. For an $A_{N-1}$ ALE space, these field strengths are supported by the compact exceptional $2$-cycles and form an $(N-1)$-dimensional lattice, in agreement with the rank of the compactly supported middle cohomology $H^2(X,\mathbb Z)$. Although the field strengths decay at large radius and the gauge fields approach flat connections at asymptotic infinity, the limiting flat bundle need not be topologically trivial. The boundary is the lens space
$
S^3/\mathbb Z_N,
$
whose fundamental group is $\mathbb Z_N$, and hence flat $U(1)$ bundles at infinity are classified by
$
\mathrm{Hom}(\mathbb Z_N,U(1))\simeq \mathbb Z_N.
$
Thus, the Maxwell fields become locally flat near infinity while retaining discrete global information through their asymptotic holonomy class. 

{\bf Summary of results}. We begin by evaluating the Maxwell path integral directly on an $A_{N-1}$ ALE space. 
A key outcome of this computation is that the answer is not naturally a single scalar partition function. Rather, the flux sum decomposes into theta-function blocks labeled by the finite quotient
$
P/Q\simeq \mathbb Z_N,
$
where $Q$ is the root lattice and $P$ is the weight lattice of $A_{N-1}$. Thus, the natural ALE object resulting from evaluating the path integral takes the form of a state
\begin{equation}\nonumber
|\Psi(\tau)\rangle
=
\sum_{\mu\in P/Q}
\Theta_\mu(\tau)\,|\mu\rangle,
\end{equation}
where the basis vectors $|\mu\rangle$ label the distinct boundary sectors and the functions $\Theta_\mu(\tau)$ are the corresponding ALE theta blocks.

This decomposition is one of the main points of the paper. On a compact closed manifold, the Maxwell path integral gives a number. On an ALE space, by contrast, the noncompact end supplies an asymptotic boundary, and the path integral should instead be interpreted as preparing a state in the Hilbert space associated with that boundary. The appearance of the $N$ theta functions is precisely the manifestation of this fact. The labels $\mu\in P/Q$ coincide with the possible flat $U(1)$ holonomy classes on the asymptotic lens-space boundary
$
\partial X = S^3/\mathbb Z_N.
$
Equivalently, they label the homomorphism from $H_1(S^3/\mathbb Z_N,\mathbb Z)\simeq \mathbb Z_N$ to $U(1)$. In this sense, the ALE theta functions are naturally interpreted as the components of a boundary state, resolved according to the discrete holonomy sector at infinity.

The duality properties of these blocks then take a natural vector-valued form. Under the modular generators $\mathbb S$ and $\mathbb T$, the individual functions $\Theta_\mu$ do not transform as independent modular forms. Instead, they mix among themselves:
\begin{equation}\nonumber
\Theta_\mu \longmapsto \sum_{\nu\in P/Q} M_{\mu\nu}\,\Theta_\nu,
\end{equation}
with the matrices $M_{\mu\nu}$ implementing a finite Fourier-transform action on the boundary-sector labels, which amounts to changing the electric to dual magnetic basis. Thus, the ALE path integral transforms as a modular vector, or equivalently as a boundary state with components $\Theta_\mu$, rather than as a single modular form. From this perspective, the apparent failure of modularity on ALE spaces that was reported in \cite{Bianchi:1996zj} is not a violation of $S$-duality, but a signal that the correct object is vector-valued: duality acts on the full set of holonomy sectors at infinity.

We then test this boundary-state interpretation by performing an explicit gluing calculation in the simplest case, namely the Eguchi--Hanson space \cite{Eguchi:1978gw}, which is the $A_1$ ALE space. Gluing Eguchi--Hanson space to its orientation-reversed copy along the common lens-space boundary produces a closed $4$-manifold diffeomorphic to $S^2\times S^2$. Unlike the ALE path integral, which gives a vector of boundary-sector amplitudes, the path integral on the glued manifold is an ordinary scalar partition function. In this language, the gluing operation is expressed as a pairing of boundary states. If
$
|\Psi_R(\tau)\rangle
=
\sum_{\mu\in P/Q}\Theta_\mu(\tau)\,|\mu\rangle
$
is the state prepared by the right ALE space, then the orientation-reversed ALE space prepares the conjugate state
$
\langle \Psi_L(\bar\tau)|
=
\sum_{\mu\in P/Q}\overline{\Theta_\mu(\tau)}\,\langle \mu| .
$
Using the natural pairing of boundary sectors,
$
\langle \mu|\nu\rangle=\delta_{\mu\nu},
$
the glued partition function is
\begin{equation}\nonumber
Z_{\rm glue}(\tau,\bar\tau)
=
\langle \Psi_L(\bar\tau)|\Psi_R(\tau)\rangle
=
\sum_{\mu\in P/Q}
\overline{\Theta_\mu(\tau)}\,\Theta_\mu(\tau).
\end{equation}
For Eguchi--Hanson, where $P/Q\simeq\mathbb Z_2$, this pairing gives precisely the Maxwell partition function on  $S^2\times S^2$.

The vector-valued modular behavior of the ALE theta blocks and the gluing procedure are also reminiscent of the modular transformation of conformal blocks in $2$-dimensional conformal field theory (CFT) \cite{DiFrancesco:1997nk}. In a rational CFT, individual chiral blocks are not modular invariant by themselves; rather, they transform among one another under the modular group, and a modular-invariant partition function is obtained only after pairing left- and right-moving sectors. The ALE construction exhibits an analogous structure. This analogy is not meant to identify the $4$-dimensional Maxwell theory with a $2$-dimensional CFT, but rather to emphasize the common structural feature.

Maxwell theory also possesses electric and magnetic $1$-form global symmetries, and the ALE construction can be refined by coupling these symmetries to background fields. We therefore study the ALE theta blocks in the presence of electric and magnetic $1$-form symmetry backgrounds and track their transformation under the modular generators $\mathbb S$ and $\mathbb T$. In this refined setting, duality not only mixes the theta-function components labeled by the boundary holonomy sectors, but also rotates the background fields themselves. Moreover, the theta blocks acquire background-dependent phases under large background gauge transformations. Equivalently, in the presence of $1$-form backgrounds the ALE blocks should not be regarded as ordinary functions on the background torus, but as sections of a line bundle over it. This line-bundle structure is the background-field manifestation of the mixed electric-magnetic $1$-form anomaly. We then repeat the Eguchi--Hanson gluing analysis with these backgrounds turned on and show that the resulting expression agrees with the Maxwell partition function on $S^2\times S^2$ in the corresponding electric and magnetic background sectors.

We further consider discrete gaugings of the $1$-form symmetries. After restricting the continuous electric or magnetic $U(1)$ $1$-form symmetry to a finite subgroup $\mathbb Z_k$, with $k$ odd, we construct gauged ALE boundary states by summing over the corresponding discrete $2$-form background sectors. These sums may be weighted by quadratic refinements, which play the role of discrete theta angles, as well as by Fourier characters selecting definite dual sectors. This leads to a family of gauged boundary-state amplitudes, including pure electric, pure magnetic, and dyonic gaugings. We then examine how these gauged states behave under Eguchi--Hanson gluing. For odd $k$, the discrete background sectors on the two ALE halves give a complete parametrization of the $\mathbb Z_k$ background sectors on the glued manifold, and the gluing reproduces the expected $\mathbb Z_k$-gauged Maxwell partition function on $S^2\times S^2$.

{\bf Organization of the paper.} The paper is organized as follows. In Section \ref{Type-A ALE space}, we review the geometry of $A$-type ALE spaces, emphasizing the compact exceptional cycles, their intersection form, and the normalizable harmonic $2$-forms that support Maxwell fluxes. We also explain how these field strengths become flat at asymptotic infinity while retaining discrete holonomy data on the lens-space boundary.
In Section \ref{Maxwell theory and S-duality}, we study Maxwell theory and its $S$-duality properties. We first recall the compact benchmark of $S^2\times S^2$, where the Maxwell path integral is a partition function with well-defined modular behavior. We then turn to ALE spaces and show that the Maxwell path integral decomposes into theta-function blocks labeled by the finite quotient $P/Q$. 
In Section \ref{Gluing ALE spaces and modularity}, we test this interpretation by gluing Eguchi--Hanson space to its orientation-reversed copy. The resulting closed manifold is diffeomorphic to $S^2\times S^2$, and we show that the natural pairing of the two ALE boundary states reproduces the Maxwell partition function on $S^2\times S^2$. 
In Section \ref{Electric and magnetic 1-form symmetry backgrounds and their gauging}, we refine the analysis by turning on electric and magnetic $1$-form symmetry backgrounds. We derive the modular transformation laws of the ALE theta blocks in the presence of these backgrounds, including the background-dependent phases, and then repeat the gluing analysis. We also study discrete $\mathbb Z_k$ gaugings of the $1$-form symmetries on Eguchi--Hanson space, including pure electric, pure magnetic, and dyonic gaugings. Finally, we examine how these gauged boundary states glue, showing that for odd $k$ the gluing reproduces the expected $\mathbb Z_k$-gauged partition function on $S^2\times S^2$.
 Appendix \ref{Sduality on the path integral} collects technical details on the modular transformations' action on the path integral defined on a manifold with boundary. Appendix \ref{app:EH-cohomology-backgrounds} explains the basic homology and cohomology of the Eguchi--Hanson space.

 {\bf In lieu of a conclusion.}  It is useful to highlight the role of locality and unitarity, or more precisely reflection positivity, which underlie the entire construction. The Eguchi--Hanson example provides a concrete realization of these principles. Locality implies that the path integral on a manifold with boundary does not directly produce a number, but rather a state in the Hilbert space associated with the boundary sectors. Gluing then corresponds to pairing such states by summing over compatible boundary data. In the present setting, gluing Eguchi--Hanson space to its orientation reversal reproduces the Maxwell partition function on the closed manifold $S^2\times S^2$. Additionally, $S^2\times S^2$ is reflection-positive since it admits an orientation-reversing involution whose fixed locus separates it into two mirror Eguchi--Hanson halves. The orientation-reversed half therefore prepares the dual state, and the partition function on the closed manifold is obtained as the corresponding Hilbert-space pairing. In this way, the gluing construction gives an explicit realization of locality, while reflection positivity becomes manifest only after passing to the reflected closed geometry.

Turning on electric and magnetic $1$-form symmetry backgrounds refines this picture further. The ALE theta blocks are then not ordinary functions on the background torus, but sections of a line bundle, reflecting the mixed electric-magnetic $1$-form anomaly. The gluing computation shows that these relative background data combine correctly into the standard background-dependent Maxwell partition function on $S^2\times S^2$. Thus the anomaly is already visible at the level of the boundary-state amplitudes, before the final gluing is performed.
We also considered discrete $\mathbb Z_k$ gaugings of the $1$-form symmetries. For odd $k$, the gluing of gauged Eguchi--Hanson blocks reproduces the expected closed-manifold $\mathbb Z_k$-gauged partition function. 

Taken together, these results suggest a broader lesson. On noncompact manifolds, and more generally on manifolds with boundary, the natural observables of quantum field theory are often sector-resolved boundary amplitudes rather than scalar partition functions. Duality, anomaly, and gauging are then most naturally formulated at this relative level, while the familiar closed-manifold partition function emerges only after the appropriate gluing. In this sense, ALE spaces furnish a broader arena for exploring locality, reflection positivity, generalized symmetries, anomalies, and duality, while making visible structures of quantum field theory that remain hidden on closed manifolds.

\section{$A$-type ALE space}
\label{Type-A ALE space}

\begin{figure}[t]
    \centering
 \includegraphics[height=0.5\textheight,angle=-90]{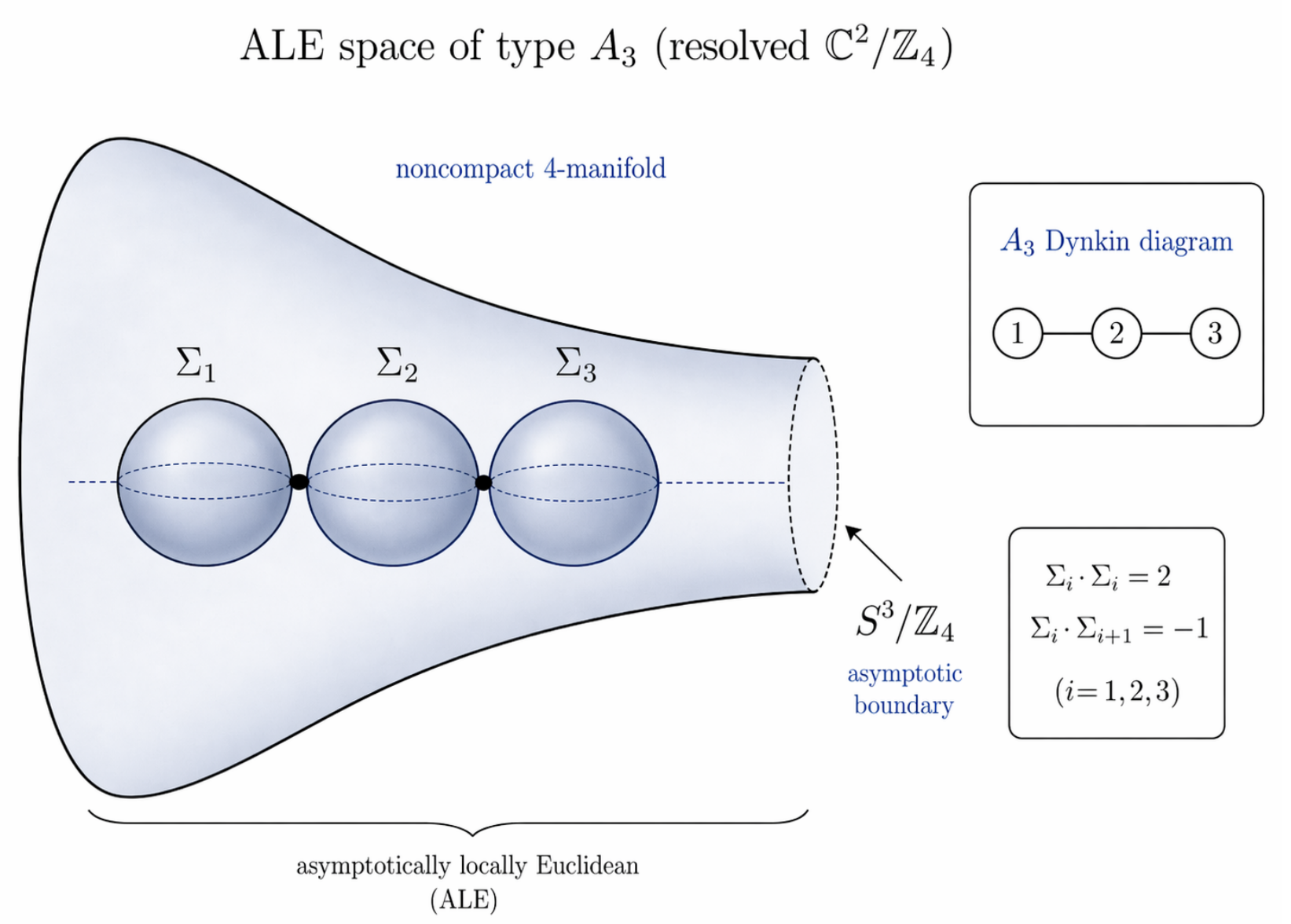}
\caption{Schematic illustration of an ALE space of $A_3$-type, equivalently the resolution of the orbifold singularity $\mathbb C^2/\mathbb Z_4$. There is a chain of three $2$-spheres $\Sigma_1,\Sigma_2,\Sigma_3$, whose intersections reproduce the $A_3$ Dynkin diagram. The asymptotic boundary is the lens space $S^3/\mathbb Z_4$. In the convention shown, the intersection pairing is written as the positive Cartan matrix, with $\Sigma_i\cdot\Sigma_i=2$ and $\Sigma_i\cdot\Sigma_{i\pm1}=-1$.}
\label{fig:A3-ALE}
\end{figure}

We begin with the Gibbons--Hawking multi-center metric \cite{Gibbons:1978tef}, which furnishes a standard realization of the ALE space of $A_{N-1}$-type. Let $\bm{x}\in\mathbb{R}^3$ and let $t$ denote the coordinate along the $S^1$ fiber. The metric is
\begin{eqnarray}
ds^2 = V^{-1}(\bm{x})\bigl(dt+\bm{{\cal W}}\cdot d\bm{x}\bigr)^2
      + V(\bm{x})\, d\bm{x}\cdot d\bm{x}\,,
\end{eqnarray}
where
\begin{eqnarray}\label{defV}
V(\bm{x})=\sum_{\mu=1}^{N}\frac{1}{|\bm{x}-\bm{x}_\mu|}
\equiv \sum_{\mu=1}^{N}V^{(\mu)}(\bm{x})
\end{eqnarray}
is a harmonic function with $N$ centers at $\bm{x}_\mu$, and the potential $1$-form ${\cal W}=\bm{{\cal W}}\cdot d\bm{x}$ satisfies
$
d{\cal W} = \star_3 dV\,,
$
or equivalently $\nabla\times\bm{{\cal W}}= \nabla V$, where $\star_3$ is the Hodge star with respect to the flat metric on $\mathbb R^3$. This condition ensures that the resulting spin connection $1$-form is self-dual,
$
\tilde{\omega}^{a}{}_{b}
= \frac{1}{2}\epsilon_{abcd}\,\omega^{c}{}_{d}
=  \omega^{a}{}_{b}\,,
$
and hence the curvature $2$-form obeys
$
\tilde{R}^{a}{}_{b}
= \frac{1}{2}\epsilon_{abcd}\,R^{c}{}_{d}
=  R^{a}{}_{b}\,.
$
It follows that the Gibbons--Hawking space is Ricci-flat and solves the vacuum Einstein equations, $R_{ab}=0$. In this sense, the ALE space is a gravitational instanton. 

The ALE space admits $N-1$ compact $2$-cycles $\{\Sigma_{i}\}$, $i=1,2,..,N-1$, obtained by fibering the $S^1$ circle over intervals joining neighboring centers\footnote{For instance, fibering the $S^1$ over an interval such as $[\bm x_1,\bm x_2]$ or $[\bm x_2,\bm x_3]$ produces a compact $2$-cycle: the circle fiber is nontrivial over the interior of the interval and shrinks to a point at both endpoints, so the space is topologically $S^2$.}. 
These $2$-cycles  generate the homology $H_2(X,\mathbb Z)$ and may be identified with the simple roots $\{\bm \alpha_i\}$ of the root lattice $Q$ of the $A_{N-1}$ algebra. The $2$-cycle intersection form is\footnote{Our definition of the intersection form differs by a negative sign from that in the mathematics literature, see, e.g., \cite{Donaldson1986Connections,Kronheimer:1989zs}. Thus, the ALE instantons in our convention are self-dual rather than anti-self-dual.}
\begin{eqnarray}
\Sigma_i\cdot \Sigma_j=\,C_{ij}\,,
\end{eqnarray}
where $C_{ij}$ is the Cartan matrix: 
\begin{eqnarray}
C_{ij}\equiv \bm\alpha_i\cdot\bm \alpha_j=2\delta_{ij}-\delta_{i,j-1}-\delta_{i,j+1}\,,\quad i,j=1,2,..,N-1\,.
\end{eqnarray}
Here, we follow the convention that the roots have length $\bm \alpha^2=2$, and hence, the root lattice coincides with the dual root lattice, i.e., $\bm\alpha^*\equiv \frac{2}{\bm\alpha^2}\bm\alpha=\bm\alpha$. See Figure \ref{fig:A3-ALE} for a cartoonish description of the ALE space associated with the $A_3$ algebra. 

$A$-type ALE space has $N-1$ normalizable harmonic self-dual $2$-forms dual to the $2$-cycles, which are naturally valued in the dual lattice, namely the weight lattice $P$ of the $A_{N-1}$ algebra. More precisely, one may choose a basis of cohomology classes $\bigl\{[F_i/2\pi]\bigr\}\subset H^2(X,\mathbb Z)$ dual to the cycles $\Sigma_i$, so that
\begin{eqnarray}\label{cohmoclass}
\int_{\Sigma_i}\frac{F_j}{2\pi}=\delta_{ij}\,.
\end{eqnarray}
We identified this basis with the fundamental-weight basis of the $A_{N-1}$ weight lattice, writing schematically
$
F_i\equiv 2\pi\,\bm w_i,
$
where $\{\bm w_i\}$, $i=1,2,..,N-1$ is the set of the fundamental weights. 
The pairing on the dual lattice (weight lattice) is given by the inverse of the intersection form, and one finds\footnote{\label{footnoteC}To obtain this result, we write schematically $F_i=2\pi\bm w_i$. Then, $F_i\cdot F_j=(2\pi)^2\bm w_i\cdot \bm w_j=(2\pi)^2\left(\mbox{min}(i,j)-\frac{ij}{N}\right)=(2\pi)^2(C^{-1})_{ij}$. The latter result can be obtained by writing $\bm w_i=\sum_{a=1}^i\bm e_a-\frac{i}{N}\sum_{a=1}^N\bm e_a$, and $\{\bm e_a\}$, $a=1,2,..,N$, are the standard unit vectors on $\mathbb R^N$.}
\begin{eqnarray}\label{inverseC}
\frac{1}{8\pi^2}\int F_i\wedge F_j
=
\frac{1}{2}\,(C^{-1})_{ij}=\frac{1}{2}\left(\mbox{min}(i,j)-\frac{ij}{N}\right)\,.
\end{eqnarray}
More generally, a general self-dual $2$-form  $F$ can be decomposed as
\begin{eqnarray}\label{sumF}
F=\sum_{i=1}^{N-1} m_i\,F_i\,,
\end{eqnarray}
with $m_i\in\mathbb Z$. Then,  one obtains\footnote{We shall identify the self-dual $2$-form $F$ with the $U(1)$ field strength, and then, the quantity $\frac{1}{8\pi^2}\int F\wedge F$ can be identified as the topological charge of the $U(1)$ field.}
\begin{eqnarray}\label{mainintsform}
\frac{1}{8\pi^2}\int F\wedge F
=
\frac{1}{2} \sum_{i,j=1}^{N-1}m_i\,(C^{-1})_{ij}\,m_j\,,
\end{eqnarray}
which is fractional. 

The ALE space of $A_{N-1}$  type has Euler characteristic
$
\chi = N
$
and Hirzebruch signature
$
\sigma = (N-1).$
These topological invariants have a simple geometric interpretation \cite{Anselmi:1993sm}. The resolution of $\mathbb C^2/\mathbb Z_N$ introduces a chain of $N-1$ exceptional $2$-spheres, so the second Betti number is
$
b_2=N-1.
$
Since the ALE space is connected, one has
$
b_0=1. 
$
The remaining Betti numbers are $b_1 = b_3 =b_4= 0$.
Therefore, the Euler characteristic is
$
\chi=\sum_{j=0}^4(-1)^{j}b_j=N.
$
The signature is determined by the duality type of the harmonic $2$-forms. In the orientation convention used here, the $N-1$ normalizable harmonic $2$-forms are self-dual, so
$
b_2^+=N-1,
b_2^-=0,
$
and hence
$
\sigma=b_2^+-b_2^-=N-1.
$

Provided the necessary geometric data are respected, a theory with a given internal symmetry can be placed on an ALE space. Internal symmetries, such as the charge conjugation symmetry ${\mathbb C}$, are therefore not problematic in themselves. Space reflection and time-reversal symmetries, however, are more subtle,  and we choose not to consider them in this work
\footnote{A Euclidean realization of either symmetry would require an orientation-reversing involution of the underlying manifold \cite{Witten:2025ayw}, but an ALE space with a chosen orientation does not, in general, admit such a symmetry, since it has a nonvanishing Hirzebruch signature \cite{MilnorHusemoller1973}.}.


The simplest theory one can consider on an ALE space is free Maxwell theory. A convenient ansatz for a closed $U(1)$ field strength is \cite{Bianchi:1996zj}
\begin{equation}
{\cal F}
=
E_\ell(\bm{x})
\left(
e^0\wedge e^\ell
+
\frac{1}{2}\epsilon_{\ell mn}\,e^m\wedge e^n
\right),
\label{eq:ALE-Maxwell-field-strength}
\end{equation}
where the vielbeins of the orthonormal frame are
$
e^0
=
V^{-1/2}
\left(
dt+\bm{{\cal W}}\cdot d\bm{x}
\right),
e^\ell
=
V^{1/2}dx^\ell,
$
for $\ell,m,n=1,2,3$.
Introducing the $1$-form
$
\Phi:=E_\ell\,dx^\ell
$
on the $\mathbb R^3$ base, ${\cal F}$ can be written as
$
{\cal F}
=
\left(dt+\bm{\mathcal W}\!\cdot\! d\bm x\right)\wedge \Phi
+
V\,\star_3 \Phi.
\label{eq:Fphi}
$
 In this form, the closure condition $d{\cal F}=0$  is equivalent to the pair of equations (i)
$
d\Phi=0,
$
(ii)
$
d{\mathcal W}\wedge \Phi+dV\wedge \star_3\Phi+V\,d(\star_3\Phi)=0.
\label{eq:closurepair}
$
The first equation implies locally that
$
\Phi=dH,
E_\ell=\nabla_\ell H
$
for some scalar function $H$ on the $\mathbb R^3$ base. Using $d{\cal W}=\star_3 dV$, the second equation becomes
$
\star_3 dV\wedge dH+dV\wedge\star_3 dH+V\,d(\star_3 dH)=0
$, which is equivalent to
$
\nabla\!\cdot\!\bigl(V^2\nabla H\bigr)=0.
\label{eq:V2Hharmonic}
$
This equation is solved by
$
H^{(\mu)}=V^{(\mu)}/V,
$
away from the Gibbons--Hawking centers.
 Therefore
\begin{equation}
E_\ell^{(\mu)}
=
\nabla_\ell\!\left(\frac{V^{(\mu)}}{V}\right),
\qquad
\mu=1,\ldots,N,
\end{equation}
solves the closure condition. These $N$ solutions are not independent, since
$
\sum_{\mu=1}^N V^{(\mu)}/V=1
$
and hence
$
\sum_{\mu=1}^N E_\ell^{(\mu)}
=0.
$
Thus, only $N-1$ independent field strengths remain, in agreement with the rank of the compactly supported cohomology of the $A_{N-1}$ ALE space.

The field strength ${\cal F}$ is self-dual by construction. Indeed, in the orthonormal frame, one has
$
\star\!\left(e^0\wedge e^\ell\right)
=
\frac{1}{2}\epsilon_{\ell mn}\,e^m\wedge e^n,
$
and therefore
$
\star {\cal F}={\cal F}.
$
Hence ${\cal F}$ automatically satisfies the Maxwell equation
$
d\star{\cal F}=d{\cal F}=0,
$ and thus is a harmonic $2$-form. 
Moreover, for a self-dual Abelian field in $4$ Euclidean dimensions, the Maxwell stress tensor vanishes identically, so the gauge field does not backreact on the metric. Thus, the ALE metric together with the field strength ${\cal F}$ solves the Einstein--Maxwell equations.

In our analysis, we express ${\cal F}$ in the basis of cohomology classes $F_i$ defined in \eqref{cohmoclass}. Accordingly, we identify ${\cal F}=F$ and write, as in \eqref{sumF},
\begin{equation}
{\cal F}=F=\sum_{i=1}^{N-1}m_iF_i,
\end{equation}
with $m_i\in\mathbb Z$ are the ``magnetic" numbers.

It is also useful to make the asymptotic behavior explicit. At large radius, the Gibbons--Hawking potential has the form
\begin{equation}
V(\bm x)=\frac{N}{|\bm x|}+O(|\bm x|^{-2}),
\qquad
V^{(\mu)}(\bm x)=\frac{1}{|\bm x|}+O(|\bm x|^{-2}),
\end{equation}
so that
\begin{equation}
\frac{V^{(\mu)}}{V}
=
\frac1N+O(|\bm x|^{-1}),
\qquad
E_\ell^{(\mu)}
=
\nabla_\ell\!\left(\frac{V^{(\mu)}}{V}\right)
=
O(|\bm x|^{-2}).
\end{equation}
Since
$
e^0\wedge e^\ell
=
\left(dt+\bm{\mathcal W}\!\cdot\! d\bm x\right)\wedge dx^\ell,\,
e^m\wedge e^n
=
V\,dx^m\wedge dx^n,
$
the two terms in \eqref{eq:ALE-Maxwell-field-strength} decay respectively as $O(|\bm x|^{-2})$ and $O(|\bm x|^{-3})$. Thus
\begin{equation}
{\cal F}=O(|\bm x|^{-2})
\end{equation}
and in particular
\begin{equation}
{\cal F}\to 0
\qquad
\text{as } |\bm x|\to\infty.
\end{equation}
Therefore the connection becomes asymptotically flat. More precisely, on the asymptotic lens-space boundary $S^3/\mathbb Z_N$, the restriction of the field strength vanishes,
\begin{equation}
{\cal F}\big|_{\partial X_\infty}=0,
\end{equation}
so the gauge field approaches a flat $U(1)$ bundle at infinity. Although the curvature vanishes asymptotically, the bundle need not be topologically trivial there. Indeed, the asymptotic boundary is the lens space $S^3/\mathbb Z_N$, whose first homology is torsion\footnote{Starting from the Gibbons--Hawking metric,
one sees the lens-space boundary by taking the large-radius limit. For $r=|\bm x|\to\infty$,
$
V\sim \frac{N}{r},
$
and $\bm{{\cal W}}$ approaches the Dirac monopole connection of charge $N$ on the asymptotic $S^2$. Thus, up to gauge convention,
$
\bm{{\cal W}}\sim N\cos\theta\,d\phi.
$
The metric, therefore, becomes
$
ds^2
\sim
\frac{r}{N}(dt+N\cos\theta\,d\phi)^2
+
\frac{N}{r}\left(dr^2+r^2d\Omega_2^2\right).
$
Introducing $\rho=2\sqrt{Nr}$ gives
$
ds^2
\sim
d\rho^2
+
\frac{\rho^2}{4}
\left[
d\theta^2+\sin^2\theta\,d\phi^2
+
\left(\frac{dt}{N}+\cos\theta\,d\phi\right)^2
\right].
$
The expression in brackets is the metric on $S^3$ written as a Hopf fibration, with fiber coordinate $\psi=t/N$. Since $t$ has period $4\pi$, the coordinate $\psi$ has period $4\pi/N$, rather than the usual $4\pi$. Hence, the asymptotic $3$-manifold is the lens space
$
S^3/\mathbb Z_N.
$
},
\begin{equation}
H_1(S^3/\mathbb Z_N,\mathbb Z)\cong \mathbb Z_N,
\end{equation}
and since $\mbox{Hom}\left(\mathbb Z_N,U(1)\right)\cong \mathbb Z_N$, the flat $U(1)$ bundles on the boundary are classified by
$
\mathbb Z_N.
$
Equivalently, the possible asymptotic holonomies of the Maxwell connection fall into $\mathbb Z_N$ classes. Thus ${\cal F}$ becomes flat at infinity, while still retaining nontrivial global information through the discrete holonomy sector of the asymptotic flat bundle.

\section{Maxwell theory and $S$-duality}
\label{Maxwell theory and S-duality}

We consider free Maxwell theory in four dimensions, supplemented by the topological $\theta$-term,
\begin{equation}
S
=
\frac{1}{g^{2}}\int F\wedge \star F
-
\frac{i\theta}{8\pi^{2}}\int F\wedge F\,.
\end{equation}
It is convenient to combine the gauge coupling $g$ and the theta angle $\theta$ into the complexified coupling
\begin{equation}
\tau
\equiv
\frac{\theta}{2\pi}
+
i\,\frac{4\pi}{g^{2}}
\in\mathbb H\,,
\end{equation}
where
\begin{equation}
\mathbb H
=
\left\{
\tau\in\mathbb C
\ \middle|\
\operatorname{Im}\tau>0
\right\}
\end{equation}
denotes the upper half-plane.

Free Maxwell theory admits an $SL(2,\mathbb Z)$ duality action. At the level of the complexified coupling, this action is
\begin{equation}
\tau
\longmapsto
\frac{a\tau+b}{c\tau+d}\,,
\qquad
a,b,c,d\in\mathbb Z\,,
\qquad
ad-bc=1\,.
\end{equation}
 The standard matrix generators of the $SL(2,\mathbb Z)$ duality are
\begin{equation}
S_0
=
\begin{pmatrix}
0 & -1\\
1 & 0
\end{pmatrix},
\qquad
T_0
=
\begin{pmatrix}
1 & 1\\
0 & 1
\end{pmatrix},\label{fundS0T0}
\end{equation}
and induce the transformations
\begin{equation}
S_0:\quad
\tau\longmapsto-\frac{1}{\tau}\,,
\qquad
T_0:\quad
\tau\longmapsto\tau+1\,.
\end{equation}
As elements of $SL(2,\mathbb Z)$, these matrices satisfy
\begin{equation}
S_0^{4}=I\,,
\qquad
S_0^{2}=(S_0T_0)^{3}=-I\,,
\qquad
(S_0T_0^{-1})^{3}=I\,.\label{S0TOCUB}
\end{equation}

It is important to distinguish the standard matrices $S_0,T_0\in SL(2,\mathbb Z)$ from the active duality operations $\mathbb S,\mathbb T$ acting on the fields, line operators, background fields, and theta-vector. The reason is that the fractional-linear action on $\tau$ is insensitive to the central element $-I$:
$
(-I)\cdot\tau
=
\frac{-\tau}{-1}
=
\tau\,.
$
Consequently, the action on the coupling factors through
$
PSL(2,\mathbb Z)
=
SL(2,\mathbb Z)/\{I,-I\},
$
and therefore cannot distinguish a matrix $\gamma$ from the matrix $-\gamma$.
The two matrices $\gamma$ and $-\gamma$ need not, however, define the same operation on the complete theory. The central element $-I$ is represented physically by charge conjugation $\mathbb C$, which acts nontrivially on the gauge field, electric-magnetic charges, and background fields:
\begin{equation}
\mathbb C:\qquad
A\longmapsto-A\,,
\qquad
(q_e,q_m)\longmapsto(-q_e,-q_m)\,,
\qquad
(b,c)\longmapsto(-b,-c)\,.
\end{equation}
Thus, although $\gamma$ and $-\gamma$ induce the same transformation of $\tau$, their active actions on the full theory differ by charge conjugation.

For this reason, we reserve $S_0$ and $T_0$ for the ordinary $SL(2,\mathbb Z)$ matrices and use $\mathbb S$ and $\mathbb T$ for the active duality operations. When acting on the fields (and later on the $\Theta$ functions), $\mathbb S$ and $\mathbb T$ denote the corresponding lifted operations, including the modular automorphy factors and the induced transformation of the background fields. With the active conventions adopted below, this lifted action satisfies
\begin{equation}
\mathbb S^2
=
(\mathbb S\mathbb T^{-1})^3
=
\mathbb C\,,
\qquad
\mathbb S^4=1\,.
\label{STOCUB}
\end{equation}
Although $\mathbb S$ induces the same fractional-linear transformation $\tau\mapsto-1/\tau$ as $S_0$, our active convention represents $\mathbb S$ by $-S_0$ on the electromagnetic charge lattice, while $\mathbb T$ is represented by $T_0$.

Let $F_D$ denote the dual $2$-form associated with $F$. Under the action of $\mathbb S$, one has
\begin{equation}
\mathbb S:\quad
F \mapsto -F_D \equiv -\left(\frac{i\,4\pi}{g^2}\,\star F + \frac{\theta}{2\pi}\,F\right)\,,\quad
F_D \mapsto F\,.
\end{equation}
This bulk duality relation can be obtained directly starting from the path integral  \cite{Witten:1995gf}; see Appendix \ref{Sduality on the path integral} where we review the duality for a theory defined on a manifold with boundary. 
Next, denote the Wilson and ’t Hooft line operators by
$
W \equiv \exp\!\left(i\oint_c A\right),
H \equiv \exp\!\left(i\oint_c A_D\right),
$
where $F=dA$, $F_D=dA_D$, and $c$ is a closed contour. Under $\mathbb S$, these line operators transform as
\begin{equation}
\mathbb S:\quad (W,H)\mapsto (H^\dagger,W)\,.
\end{equation}
Therefore, under $\mathbb S$, a dyon carrying electric and magnetic charges $(q_e,q_m)$ transforms as
\begin{eqnarray}
\mathbb S:\quad (q_e,q_m)\mapsto(q_m,-q_e).
\end{eqnarray}
Now, we see that although $\mathbb S^2$ acts trivially on $\tau$, it acts nontrivially on the gauge field, implementing the $\mathbb Z_2^{(0)}$ $0$-form charge-conjugation symmetry $\mathbb C$:
\begin{equation}
\mathbb S^2={\mathbb C}:\quad A \mapsto -A\,.
\end{equation}

The generator $\mathbb T$ shifts the $\theta$ angle by $2\pi$, or equivalently
\begin{equation}
\mathbb T:\quad \tau \mapsto \tau+1\,.
\end{equation}
While this leaves the field strength $F$ unchanged, it shifts the dual field strength according to
\begin{equation}
\mathbb T:\quad F \mapsto F\,,
\qquad
F_D \mapsto F_D + F\,.
\end{equation}
Accordingly, the Wilson and 't Hooft line operators transform as
\begin{equation}
\mathbb T:\quad (W,H)\mapsto (W,HW)\,.
\end{equation}
Thus, under $\mathbb T$, a pure 't Hooft line acquires one unit of Wilson charge. More generally, a dyonic line of electric and magnetic charges $(q_e,q_m)$ is mapped as
\begin{equation}
\mathbb T:\quad (q_e,q_m)\mapsto (q_e+q_m,q_m)\,.
\end{equation}
This is the familiar Witten effect: shifting $\theta$ by $2\pi$ induces one unit of electric charge on a magnetic line operator. Then, it is straightforward to see that
\begin{eqnarray}
\left(\mathbb S\mathbb T^{-1}\right)^3=\mathbb S^2: \quad (q_e,q_m)\mapsto (-q_e,-q_m)=\mathbb C (q_e,q_m)\,.
\end{eqnarray}

The modular group has two distinguished fixed points in the fundamental domain  $\mathbb F$ of $SL(2,\mathbb Z)$:
\begin{equation}
\mathbb F
=
\left\{
\tau \in \mathbb H
\;:\;
|\tau|\ge 1,\;
-\frac12 \le \operatorname{Re}(\tau)\le \frac12
\right\}\,.
\end{equation}
 The first is
\begin{equation}
\tau_{\star}=i,
\end{equation}
which is fixed by the $\mathbb S$ transformation. At this point, $\mathbb S$ becomes an actual symmetry of the theory. Since
$
\mathbb S^2=\mathbb C$, and
$\mathbb C^2=1
$,
it follows that
$
\mathbb S^4=1,
$
so the symmetry generated by $\mathbb S$ is $\mathbb Z_4^{(0)}$ on the full theory. The second fixed point  is
\begin{equation}
\tau_{\star\star}=e^{i\pi/3},
\end{equation}
which is fixed by $\mathbb S\mathbb T^{-1}$, since it satisfies
$
-1/(\tau_{\star\star}-1)=\tau_{\star\star}.
$
At this point, $\mathbb S\mathbb T^{-1}$ becomes an actual symmetry of the theory\footnote{In terms of the matrices $S_0$ and $T_0$ introduced in (\ref{fundS0T0}), the two matrices $S_0T_0^{-1}$ and $-S_0T_0^{-1}$ induce the same Möbius transformation,
$
\tau
\longmapsto
-\frac{1}{\tau-1}\,,
$
which fixes
$
\tau_{\star\star}
=
e^{i\pi/3}\,.
$
They nevertheless have different orders in $SL(2,\mathbb Z)$:
\begin{equation}
(S_0T_0^{-1})^{3}=I\,,
\qquad
(-S_0T_0^{-1})^{3}=-I\,.
\end{equation}
Thus, $S_0T_0^{-1}$ is the order-three representative of this transformation, whereas $-S_0T_0^{-1}$ is its order-six representative.}. Using
$
(\mathbb S\mathbb T^{-1})^3=\mathbb C$, and
$\mathbb C^2=1,
$
one finds
$
(\mathbb S\mathbb T^{-1})^6=1,
$
so the symmetry generated by $\mathbb S\mathbb T^{-1}$ is $\mathbb Z_6^{(0)}$ on the full theory.\footnote{Although the $\mathbb Z_4$ symmetry at $\tau_\star=i$ and the $\mathbb Z_6$ symmetry at $\tau_{\star\star}=e^{i\pi/3}$ are both $0$-form symmetries, they nevertheless act nontrivially on Wilson, 't Hooft, and more generally dyonic line operators. This does not make them higher-form symmetries. The point is that these fixed-point symmetries are ordinary duality transformations: they act by rotating the electric-magnetic charge lattice and hence map one line operator to another.} 

In addition, free Maxwell theory admits electric and magnetic $1$-form global symmetries,
\begin{eqnarray}
U(1)^{[1]}_e \times U(1)^{[1]}_m.
\end{eqnarray}
The electric symmetry is generated by the conserved current $\star F$, while the magnetic symmetry is generated by the conserved current $F$. Wilson loops are charged under $U(1)^{[1]}_e$, and 't Hooft loops are charged under $U(1)^{[1]}_m$. More generally, dyonic line operators carry charges under both $1$-form symmetries.

The path integral defined for Maxwell theory on a manifold $M$ should be understood as a sum over disconnected topological sectors, together with an ordinary functional integral over gauge-field fluctuations within each sector. Schematically, one may write
\begin{equation}
\mathrm{PI}_M(\tau,\bar\tau)
=
\sum_{n}
\int_{\mathcal A_n/\mathcal G}
DA\,
e^{-S[A]} .
\end{equation}
Here $n$ labels the topological sector of the gauge field, for example, its fluxes through nontrivial $2$-cycles of $M$. For fixed $n$, the integral is over gauge connections in that sector, modulo gauge transformations ${\cal G}$. Thus, the full path integral receives contributions both from the discrete sum over topological flux sectors and from the continuous fluctuations of the gauge field around each such sector.

If $M$ is closed, this path integral defines the partition function $Z$ of free Maxwell theory on a closed $4$-manifold. Examples include $\mathbb T^4$, $S^2\times S^2$, $K3$, and $\mathbb{CP}^2$. This partition function is not generally a modular-invariant scalar. Rather, under electric-magnetic duality, it transforms as a modular form whose weights are determined by the topology of $M$ \cite{Verlinde:1995mz,Witten:1995gf}.  In particular, under an $SL(2,\mathbb Z)$ transformation
$
\tau \mapsto \frac{a\tau+b}{c\tau+d}\,,
$
the partition function acquires a nontrivial prefactor. One has
\begin{equation}
Z\!\left(\frac{a\tau+b}{c\tau+d},\frac{a\bar\tau+b}{c\bar\tau+d}\right)
=
(c\tau+d)^{u}(c\bar\tau+d)^{v}
\,Z(\tau,\bar\tau)\,,
\end{equation}
up to possible $(\tau,\bar\tau)$-independent numbers. Thus, on a general closed $4$-manifold, the Maxwell partition function is modular covariant rather than modular invariant. The modular weights $(u,v)$ are controlled by the Euler characteristic $\chi$ and Hirzebruch signature $\sigma$, and therefore encode basic topological data of the underlying manifold. In particular\footnote{The modular weights can also be thought of as a mixed anomaly between $SL(2,\mathbb Z)$ and gravity \cite{Seiberg:2018ntt}.}
\begin{equation}\label{uvdata}
(u,v)=\left(\frac{\chi+\sigma}{4}, \frac{\chi-\sigma}{4}\right)\,.
\end{equation}
 We also note that the partition function is invariant under $\tau\rightarrow \tau+1$ when the manifold is spin, and under $\tau\rightarrow \tau+2$ for nonspin manifolds. 

\subsection{The partition function and modularity on $S^2\times S^2$}

A particularly simple closed $4$-manifold that is important for future reference is $S^2\times S^2$. This manifold has
$
H_2(S^2\times S^2,\mathbb Z)\cong \mathbb Z\oplus \mathbb Z,
$
generated by the two factors
$
\Sigma_1=S^2\times\{\mathrm{pt}\},
\Sigma_2=\{\mathrm{pt}\}\times S^2.
$
The intersection form is
$
\Sigma_1\cdot \Sigma_1=0,
\Sigma_2\cdot \Sigma_2=0,
\Sigma_1\cdot \Sigma_2=1.
$
Thus, a $U(1)$ flux sector is labelled by two integers,
\begin{eqnarray}\label{integralfluxs2}
\frac{1}{2\pi}\int_{\Sigma_1}F=q\,,\quad
\frac{1}{2\pi}\int_{\Sigma_2}F=p,
\end{eqnarray}
with topological charge and kinetic terms (we choose the two spheres to have equal unit volumes)
\begin{equation}
\frac{1}{8\pi^2}\int F\wedge F=pq \,,\quad\int F \wedge \star F=4\pi^2(p^2+q^2)\,.
\end{equation}
 Then, the Maxwell action in the flux sector $(p,q)$ is
\begin{eqnarray}\label{actionS2}
S
=
\frac{4\pi^2}{g^2}(p^2+q^2)
-
i\theta\,pq .
\end{eqnarray}
Ignoring the Gaussian-integral contribution coming from the quantum fluctuations, for now, i.e., caring only about the classical part, the flux contribution to the partition function is
\begin{eqnarray}\label{ZS2S2}
Z^{\rm cl}_{S^2\times S^2}(\tau,\bar \tau)
&=&\sum_{p,q\in\mathbb Z}
\exp\left[
-\frac{4\pi^2}{g^2}(p^2+q^2)
+
i\theta pq
\right].
\end{eqnarray}
Then, it can be shown via Poisson's resummation that this partition function transforms under $\mathbb S$ as a modular form:
\begin{eqnarray}\label{S2S2duality}
\mathbb S:\quad Z^{\rm cl}_{S^2\times S^2}(\tau,\bar \tau)\mapsto Z^{\rm cl}_{S^2\times S^2}(-1/\tau,-1/\bar \tau)= |\tau|Z^{\rm cl}_{S^2\times S^2}(\tau,\bar \tau)\,.
\end{eqnarray}

On a compact closed manifold, the $1$-loop factor of free Maxwell theory is obtained by gauge fixing, introducing the Faddeev--Popov ghosts, and performing the Gaussian integral over the nonzero modes. This gives, schematically (the reader is referred to \cite{Witten:1995gf} for a detailed explanation),
\begin{equation}\label{1loopdet}
Z^{\rm 1\mbox{-}loop}
\;\propto\;
g^{\,b_0-b_1}\,
\frac{\det{}'(\Delta_0)}{\det{}'(\Delta_1)^{1/2}},
\end{equation}
where $\Delta_0$ and $\Delta_1$ are the scalar and $1$-form Laplacians, respectively, and the prime indicates omission of zero modes. Thus, the topological dependence of the overall power of the coupling is governed by the zero modes of the $0$- and $1$-form Laplacians, namely the Betti numbers\footnote{We also give the topological data of $S^2\times S^2$ for convenience. The Betti numbers are
$
b_0=1, b_1=0, b_2=2, b_3=0, b_4=1.
$
Thus
$
\chi=\sum_{j=0}^4(-1)^{j}b_j=4.
$
The intersection form is
$
Q=
\begin{pmatrix}
0&1\\
1&0
\end{pmatrix},
$
which has one positive and one negative eigenvalue. Therefore
$
b_2^+=b_2^-=1$ and
$
\sigma=b_2^+-b_2^-=0.
$}
 $b_0$ and $b_1$. For $S^2\times S^2$ we have $b_0=1, b_1=0$ and thus, the $1$-loop determinant gives a factor of $g\sim 1/\sqrt{\mbox{Im}(\tau)}$, and the full partition function, modulo a $(\tau,\bar\tau)$-independent number, is
\begin{eqnarray}
Z_{S^2\times S^2}(\tau,\bar \tau)=\frac{1}{\sqrt{\mbox{Im}(\tau)}}Z^{\rm cl}_{S^2\times S^2}(\tau,\bar \tau)\,.
\end{eqnarray}
Under $\mathbb S$ transformation we obtain: $\mbox{Im}(-1/\tau)=\mbox{Im}(\tau)/\tau\bar\tau$, and thus, the full partition function  transforms under $\mathbb S$ as
\begin{eqnarray}\label{S2S2dualityfull}
\mathbb S:\quad Z^{}_{S^2\times S^2}(\tau,\bar \tau)\mapsto  \tau \bar\tau Z^{}_{S^2\times S^2}(\tau,\bar \tau)\,.
\end{eqnarray}
The modular weights are exactly those obtained from (\ref{uvdata}), recalling that for $S^2\times S^2$ we have $\chi=4$ and $\sigma=0$.

\subsection{$S$-duality on ALE space}

On an ALE space, the local classical fields and line operators transform under $\mathbb S$ and $\mathbb T$ in the same way as on a closed manifold. The local bulk theory still admits the electric and magnetic $1$-form symmetries generated by $d\star F=0$ and $dF=0$, so that Wilson and 't Hooft loops transform locally exactly as on a closed $4$-manifold. Globally, however, the situation is modified by the noncompact asymptotic boundary: topological data at the asymptotic boundary falls into $N$ discrete sectors, as we pointed out in Section \ref{Type-A ALE space}.

Next, we consider the path integral of the free Maxwell theory on ALE space. Using (\ref{sumF}, \ref{mainintsform}) along with the fact that $F$ is self-dual, the action of the free Abelian theory is given by
\begin{eqnarray}
S=-i\pi\tau\sum_{i,j=1}^{N-1}\,m_i\,(C^{-1})_{ij}\,m_j\,.
\end{eqnarray}
 Thus, the resulting ``partition function" reads
\begin{eqnarray}\label{mainPF}
Z_{\rm ALE}(\tau)=\frac{1}{{\cal C}(g)}\sum_{\{m_i\}\in \mathbb Z}\exp\left[i\pi \tau\sum_{i,j=1}^{N-1}\,m_i\,(C^{-1})_{ij}\,m_j \right]\,.
\end{eqnarray}
Here, $1/{\cal C}(g)$ is a pre-factor resulting from the $1$-loop determinant. 
On a non-compact space, the $g$-factor dependency should be determined from the normalizable zero modes of the $0$- and $1$-form Laplacians. In the case of ALE space, there are no such normalizable zero modes, so the factor $g^{\,b_0-b_1}$ from the determinants in (\ref{1loopdet}) is absent. Accordingly, one may adopt a regularization scheme in which the Gaussian determinant contributes no residual overall power of $g$, and the nontrivial coupling dependence is then carried entirely by the classical flux sum. Therefore, we set
 \begin{equation}
 C(g)=1\,.
 \end{equation}

 Notice the quotation marks we used to designate the above quantity as a partition function. Since the ALE space is noncompact and has asymptotic boundary $S^3/\mathbb Z_N$ at infinity, the Euclidean path integral is more properly interpreted as defining a state in the Hilbert space $\mathcal H_{S^3/\mathbb Z_N}$ rather than directly a complex number; we shall come to this point later in detail, which is one of the hallmarks of this work. 

Unlike the partition function of Maxwell theory on closed $4$-manifolds,  $Z_{\rm ALE}(\tau)$ is not covariant under the $SL(2,\mathbb Z)$ transformation. In particular, under $\mathbb S$ transformation, one finds
\begin{eqnarray}
Z_{\rm ALE}(-1/\tau)\neq \tau^{u} Z_{\rm ALE}(\tau)\,,
\end{eqnarray}
as noted in \cite{Bianchi:1996zj}. This difference can be traced to the structure of the underlying homology lattice. For a closed $4$-manifold, the free part of $H_2$ is equipped with a self-dual intersection form. In $A$-type  ALE space, however, the compact $2$-cycles span the root lattice of the $A_{N-1}$ Lie algebra, which is not self-dual; its dual is the weight lattice. Therefore, under $S$-duality, the partition function is naturally related to that of the dual weight lattice rather than mapping back to itself, and the naive modular invariance is lost.

To clarify the action of $S$-duality on $Z_{\rm ALE}$, it is useful to organize the sum in (\ref{mainPF}) as sums over the weight lattice according to $N$-ality sectors. Let $\lambda$ be a general element of the weight lattice $P$. Since $P$ is generated by the fundamental weights $\{\bm w_i\}$, with $i=1,\dots,N-1$, any such vector may be written as
\begin{equation}
\lambda=\sum_{i=1}^{N-1} m_i \bm w_i\,,
\qquad m_i\in\mathbb Z\,.
\end{equation}
Its square norm is therefore
\begin{equation}
\lambda\cdot\lambda
=
\sum_{i,j=1}^{N-1} m_i (C^{-1})_{ij} m_j\,,
\end{equation}
where $C^{-1}$ is the inverse Cartan matrix; see \eqref{inverseC} and Footnote~\ref{footnoteC}.

\begin{figure}[t]
    \centering
 \includegraphics[height=0.55\textheight,angle=-90]{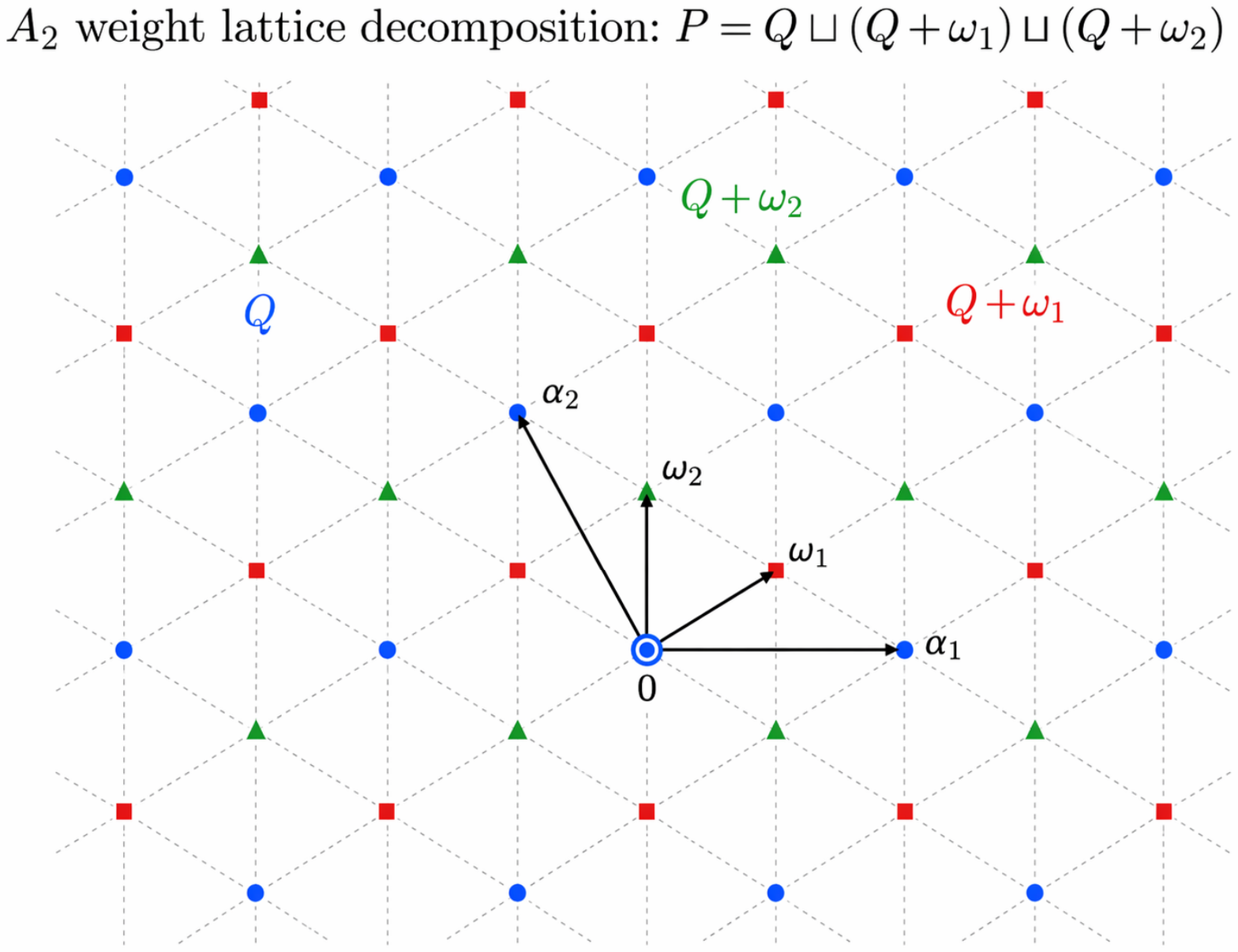}
\caption{Decomposition of the weight lattice $P$ of $A_{2}$ into cosets of the root lattice $Q$. The lattice points split into three disjoint families, $Q+\bm w_{\mu}$ with $\mu=0,1,2$, corresponding to the three elements of $P/Q\cong \mathbb Z_{3}$. The blue circles, red squares, and green triangles represent the three cosets $Q+\bm w_{0}=Q$, $Q+\bm w_{1}$, and $Q+\bm w_{2}$, respectively. Each coset defines a sector function $\Theta_{\mu}(\tau)=\sum_{\lambda\in Q+\bm w_{\mu}} e^{i\pi\tau \lambda^{2}}$. Here $\bm w_{0}=0$, while $\bm \alpha_{1}$ and $\bm\alpha_{2}$ denote the simple roots generating $Q$.}
    \label{fig:cosetlattice}
\end{figure}

Each fundamental weight $\bm w_i$ carries $N$-ality $i$, where the $N$-ality is defined as the number of boxes in the corresponding Young tableau modulo $N$. By contrast, every root has vanishing $N$-ality. It follows that the $N$-ality of a general weight
$
\lambda=\sum_{i=1}^{N-1} m_i \bm w_i
$
is given by
\begin{equation}
\mu \equiv \sum_{j=1}^{N-1} j\,m_j \pmod N\,,
\qquad
\mu=0,1,\dots,N-1\,.
\end{equation}
Using this decomposition, the partition function on the ALE space may be organized as a sum over the $N$ distinct $N$-ality sectors:
\begin{equation}\label{Zassectors}
Z_{\rm ALE}(\tau)
=
\sum_{\lambda\in P}
\exp\!\left(i\pi\tau\,\lambda\cdot\lambda\right)
=
\sum_{\mu=0}^{N-1}\Theta_\mu(\tau)\,,
\end{equation}
where
\begin{equation}\label{Thetaimport}
\Theta_\mu(\tau)
=
\sum_{\substack{\lambda\in P\\ \sum_{j=1}^{N-1} j m_j \equiv \mu \,(\mathrm{mod}\,N)}}
\exp\!\left(i\pi\tau\,\lambda\cdot\lambda\right)\,.
\end{equation}
Introducing
\begin{equation}
q \equiv e^{2\pi i\tau}\,,
\end{equation}
this can equivalently be written as
\begin{equation}
\Theta_\mu(\tau)
=
\sum_{\lambda\in \mathcal C_\mu}
q^{\frac{1}{2}\lambda\cdot\lambda}\,,
\end{equation}
where $\mathcal C_\mu$ denotes the coset of the root lattice $Q$ with $N$-ality $\mu$. More explicitly,
\begin{equation}
\mathcal C_0 = Q,
\qquad
\mathcal C_\mu = Q+\bm w_\mu,
\qquad \mu=1,\dots,N-1.
\end{equation}
We emphasize that there is no fundamental weight $\bm w_0$; rather, the sector $\mu=0$ is represented by the zero vector and corresponds to the root lattice itself. An example of this construction is illustrated in Figure \ref{fig:cosetlattice}.

Since the $SL(2,\mathbb Z)$ action is generated by $\mathbb S$ and $\mathbb T$, it is sufficient to determine how the functions $\Theta_\mu(\tau)$ transform under these two generators. Under $\mathbb T:\tau\mapsto \tau+1$, one finds
\begin{equation}
\Theta_\mu(\tau+1)
=
\sum_{\lambda\in \mathcal C_\mu}
e^{i\pi\lambda\cdot\lambda}
q^{\frac{1}{2}\lambda\cdot\lambda}.
\end{equation}
For $\lambda$ in a fixed coset $\mathcal C_\mu$, the phase $e^{i\pi\lambda\cdot\lambda}$ depends only on the $N$-ality class. Using Footnote~\ref{footnoteC}, one obtains
\begin{equation}\label{Tdualitytrans}
\mathbb T:\quad
\Theta_\mu(\tau)\mapsto \Theta_\mu(\tau+1)
=
e^{i\pi\mu^2\left(1-\frac{1}{N}\right)}\Theta_\mu(\tau).
\end{equation}
Thus, each fixed-$N$-ality sector transforms diagonally under $\mathbb T$. Notice that the genuine periodicity of the $\theta$ angle is the smallest power of $\mathbb T$ that acts trivially on all sectors $\mu\in P/Q$. In particular, the fractional topological charge on the ALE space enlarges the $\theta$-angle periodicity to\footnote{
The enlarged periodicity can be seen already from the topological charge (\ref{inverseC}). Recalling that $(C^{-1})_{ij}
=
\min(i,j)\bigl(N-\max(i,j)\bigr)/N$,
one finds the minimal charge
$
\frac{1}{8\pi^2}\int_X F_1\wedge F_1
=
\frac12(C^{-1})_{11}
=
\frac{N-1}{2N}.
$
Thus, if $N$ is odd, $N-1$ is even and this charge has denominator $N$, while if $N$ is even, $N-1$ is odd and the denominator is $2N$. Hence the minimal $\theta$-periodicity is $2\pi N$ for odd $N$, and $2\pi(2N)$ for even $N$.
}
\begin{equation}
\theta\sim \theta+2\pi N,
\qquad
N \ \mathrm{odd},
\end{equation}
and
\begin{equation}
\theta\sim \theta+2\pi(2N),
\qquad
N \ \mathrm{even}.
\end{equation}
For Eguchi--Hanson space, where $N=2$, the minimal periodicity is
\begin{equation}
\theta\sim \theta+8\pi .
\end{equation}

We now turn to the action of $\mathbb S$. By definition,
\begin{equation}
\mathbb S:\quad
\Theta_\mu(\tau)\mapsto
\Theta_\mu\!\left(-\frac{1}{\tau}\right)
=
\sum_{\lambda\in \mathcal C_\mu}
\exp\!\left(-\frac{i\pi}{\tau}\,\lambda\cdot\lambda\right).
\end{equation}
In this form, the $\mathbb S$-transformation is not yet transparent. However, as discussed above, $S$-duality naturally exchanges the homology lattice with its dual. In the present case, the relevant lattice is the root lattice $Q$, whose dual lattice is the weight lattice $P$. The appropriate tool for making this exchange manifest is the Poisson resummation formula.

For an $r$-dimensional lattice $L$ and its dual lattice $L^\vee$, Poisson resummation gives
\begin{equation}\label{PRF}
\sum_{x\in L} f(x)
=
\frac{1}{\sqrt{|L^\vee/L|}}
\sum_{\xi\in L^\vee}\hat f(\xi),
\end{equation}
where
\begin{equation}
\hat f(\xi)
=
\int_{\mathbb R^r} d^r y\,
e^{-2\pi i\,\xi\cdot y}\,f(y)
\end{equation}
is the Fourier transform of $f$.
To apply this formula, we write $\Theta_\mu(-1/\tau)$ as a sum over the root lattice. For $\mu=0$, this is simply a sum over $Q$. For $\mu=1,\dots,N-1$, the relevant coset is $Q+\bm w_\mu$. Thus, introducing representatives
\begin{equation}
\lambda_0:=0,
\qquad
\lambda_\mu:=\bm w_\mu,
\qquad \mu=1,\dots,N-1,
\end{equation}
we may uniformly write
\begin{equation}
\Theta_\mu\!\left(-\frac{1}{\tau}\right)
=
\sum_{x\in Q}
\exp\!\left[
-\frac{i\pi}{\tau}\,(x+\lambda_\mu)\cdot(x+\lambda_\mu)
\right].
\end{equation}
We therefore identify $L=Q$ in \eqref{PRF}. Since $L^\vee=Q^\vee=P$, and for $A_{N-1}$ one has
$
P/Q \cong \mathbb Z_N,
|P/Q|=N,
$
Poisson resummation yields
\begin{equation}
\Theta_\mu\!\left(-\frac{1}{\tau}\right)
=
\frac{1}{\sqrt N}
\sum_{\xi\in P}\hat f(\xi),
\end{equation}
with
\begin{equation}
\hat f(\xi)
=
\int_{\mathbb R^{N-1}} d^{N-1}y\,
e^{-2\pi i\,\xi\cdot y}
\exp\!\left[
-\frac{i\pi}{\tau}(y+\lambda_\mu)^2
\right]=
(-i\tau)^{\frac{N-1}{2}}
e^{2\pi i\,\xi\cdot\lambda_\mu}
e^{i\pi\tau\,\xi^2}.
\end{equation}
Hence,
\begin{equation}
\Theta_\mu\!\left(-\frac{1}{\tau}\right)
=
\frac{(-i\tau)^{\frac{N-1}{2}}}{\sqrt N}
\sum_{\xi\in P}
e^{2\pi i\,\xi\cdot\lambda_\mu}
e^{i\pi\tau\,\xi^2}.
\end{equation}

Next, as we did above, we decompose the weight lattice into cosets of the root lattice:
\begin{equation}
P
=
\bigcup_{\nu=0}^{N-1}\mathcal C_\nu,
\qquad
\mathcal C_0=Q,
\qquad
\mathcal C_\nu=Q+\bm w_\nu \quad (\nu=1,\dots,N-1).
\end{equation}
Accordingly, the sum over $\xi\in P$ may be written as a sum over these $N$ sectors:
\begin{equation}
\Theta_\mu\!\left(-\frac{1}{\tau}\right)
=
\frac{(-i\tau)^{\frac{N-1}{2}}}{\sqrt N}
\sum_{\nu=0}^{N-1}
e^{2\pi i\,\lambda_\mu\cdot\lambda_\nu}
\sum_{y\in \mathcal C_\nu}
e^{i\pi\tau\,y^2}.
\end{equation}
Recognizing the inner sum as $\Theta_\nu(\tau)$, we obtain
\begin{equation}
\Theta_\mu\!\left(-\frac{1}{\tau}\right)
=
\frac{(-i\tau)^{\frac{N-1}{2}}}{\sqrt N}
\sum_{\nu=0}^{N-1}
e^{2\pi i\,\lambda_\mu\cdot\lambda_\nu}\,
\Theta_\nu(\tau).
\end{equation}

For $\mu,\nu\neq 0$, we have $\lambda_\mu=\bm w_\mu$ and thus
$
e^{2\pi i\,\lambda_\mu\cdot\lambda_\nu}
=
e^{-\,\frac{2\pi i\,\mu\nu}{N}},
$
with the same formula also valid when either $\mu=0$ or $\nu=0$, since then $\lambda_0=0$. We therefore arrive at
\begin{equation}\mathbb S:\quad
\Theta_\mu(\tau)\mapsto
\Theta_\mu\!\left(-\frac{1}{\tau}\right)
=
\frac{(-i\tau)^{\frac{N-1}{2}}}{\sqrt N}
\sum_{\nu=0}^{N-1}
e^{-\,\frac{2\pi i\,\mu\nu}{N}}\,
\Theta_\nu(\tau),\label{ThetaStrans}
\end{equation}
and we recognize that the Hirzebruch signature $N-1$ appears as the modular weight.  
Thus, although $Z_{\rm ALE}(\tau)$ given by (\ref{mainPF}) does not transform covariantly under $\mathbb S$, but instead as (using (\ref{Zassectors}, \ref{ThetaStrans}))
\begin{eqnarray}
\mathbb S: Z_{\rm ALE}(\tau)\mapsto  \sqrt N(-i\tau)^{\frac{N-1}{2}}\Theta_0(\tau)\,,
\end{eqnarray}
 the individual components $\Theta_\mu(\tau)$ transform among themselves as the components of a vector-valued modular form. In this sense, the failure of modular covariance of the full ALE ``partition function" is replaced by a vector-valued modular structure.

It is also straightforward, although somewhat tedious (which requires the use of  Gaussian sums), to verify by direct iteration of the modular transformations that
\begin{equation}
(\mathbb S\mathbb T^{-1})^3 \Theta_\mu
=
\mathbb S^2 \Theta_\mu
=
\mathbb C\,\Theta_\mu
=
\Theta_{-\mu \,(\mathrm{mod}\, N)}\,.
\end{equation}
Equivalently, for $\mu\neq 0$ this may be written as
\begin{equation}
(\mathbb S\mathbb T^{-1})^3 \Theta_\mu
=
\Theta_{N-\mu}\,,
\end{equation}
while the sector $\mu=0$ is left invariant.  This is exactly as expected from the modular group relations: $(\mathbb S\mathbb T^{-1})^3$ must act in the same way as $\mathbb S^2$, and in the present case $\mathbb S^2$ is realized as charge conjugation on the sectors labeled by $\mu$, sending theta block with $N$-ality $\mu$ to a block with $N$-ality $N-\mu$.

Since the asymptotic boundary of a $A$-type ALE space is the lens space $S^{3}/\mathbb Z_{N}$, whose first homology is torsion:
$
H_{1}(S^{3}/\mathbb Z_{N},\mathbb Z)\cong \mathbb Z_{N}
$, the corresponding asymptotic sectors are labeled by the holonomy of the gauge field around this cycle, namely by the value of the Wilson loop wrapping the nontrivial boundary torsion-cycle. In this sense, the label $\mu\in P/Q\cong \mathbb Z_{N}$ in (\ref{Zassectors}) may be understood as specifying the discrete boundary holonomy at infinity. 
Accordingly, the functions $\Theta_\mu(\tau)$  label $N$ distinct sectors of the Hilbert space associated with the discrete boundary data at infinity, with modular transformations acting by mixing these sectors. This is consistent with our earlier warning that the path integral on ALE space does not define a partition function, but rather a state on Hilbert space. We further explain this assertion in the next section.

\subsection{Boundary states and duality}

In this section, we argue that the naive composition (\ref{Zassectors}) should not be understood as a partition function on the ALE space. Rather, this composition should be regarding the expansion of a boundary state into the basis of $N$ distinct boundary bases.

\begin{figure}[t]
    \centering
 \includegraphics[height=0.4\textheight,angle=-90]{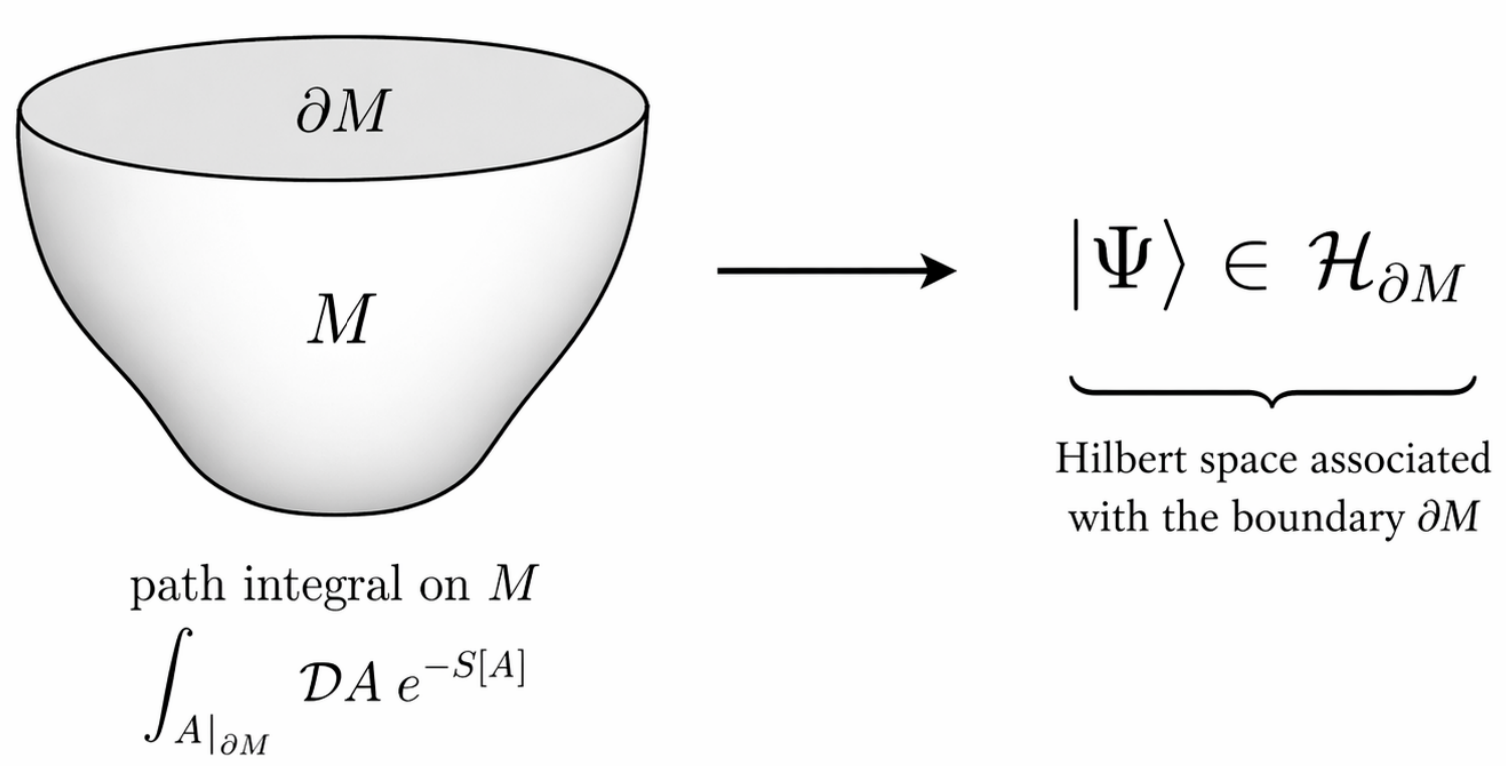}
    \caption{A path integral on a manifold $M$ with boundary $\partial M$ prepares a state $|\Psi\rangle$ in the Hilbert space $\mathcal H_{\partial M}$ associated with the boundary.}
    \label{fig:path-integral-state}
\end{figure}

If a $4$-manifold $M$ has a boundary $Y=\partial M$, the path integral should not be regarded merely as a number. Rather, it defines a state in the Hilbert space associated with the boundary (see Figure \ref{fig:path-integral-state})
\begin{equation}
\mathrm{PI}_M\in \mathcal H_Y.
\end{equation}
More specifically, in the case of free Maxwell theory, after choosing a boundary condition, for example, the Dirichlet boundary condition, in which the boundary value
\begin{equation}
\mbox{at}\, Y:\quad a=A_\parallel
\end{equation}
is fixed, this state is represented by a wavefunctional
\begin{equation}
\Psi_\tau[a]
=
\int_{A_\parallel=a}\mathcal D A\,e^{-S[A]}.
\end{equation}
In this language, electric-magnetic duality is naturally interpreted as a change of boundary polarization. Locally, and suppressing the gauge quotient and normalization factors, the $\mathbb S$-dual wavefunctional is obtained by a functional Fourier transform,
\begin{equation}
\Psi_{-1/\tau}[a_D]
=
\int \mathcal D a\,
\exp\left[
\frac{i}{2\pi}\int_Y a\wedge da_D
\right]
\Psi_\tau[a].
\label{eq:boundary-fourier-transform}
\end{equation}
Here $a_D=(A_D)_\parallel$ is the boundary value of the dual gauge field; see Appendix \ref{Sduality on the path integral} for the detailed derivation of (\ref{eq:boundary-fourier-transform}). In deriving this result, we ignored any topological degrees of freedom that might live on the boundary, as such degrees of freedom do not appear naturally in free Maxwell theory on the boundary of ALE space, the space of interest to us. The kernel in \eqref{eq:boundary-fourier-transform} is the local differential-form expression for the pairing between the boundary gauge field $a$ and its conjugate variable $a_D$. More globally, especially in the presence of torsion holonomy, this expression must be replaced by the corresponding differential-cohomology pairing.

In the ALE space, the situation is particularly simpler, since the bulk $U(1)$ bundle is restricted at infinity to a flat line bundle with nontrivial torsion holonomy. Thus, one has
$
F_\parallel\to 0
$
at infinity, while the boundary connection is not gauge-trivial. The torsion sectors are classified by
$
H_1(Y\equiv S^3/\mathbb Z_N;\mathbb Z)
\cong
\mathbb Z_N,
$
and $\mbox{Hom}(\mathbb Z_N,U(1))=\mathbb Z_N$.
In this case, the $S$-duality action, Eq. (\ref{eq:boundary-fourier-transform}), is replaced by
\begin{eqnarray}\label{SonPsi}
\Psi_{-1/\tau,\,\mu}=
\frac{1}{\sqrt N}
\sum_{\nu=0}^{N-1}
\exp\left[2\pi i\,\lambda(\mu,\nu)\right]\Psi_{\tau,\,\nu}\,,
\end{eqnarray}
where $\lambda(\mu,\nu)$ is the linking-pairing between the electric and magnetic discrete labels. For the ALE case, this pairing is given by the discrete-Fourier transform kernel: $
\lambda(\mu,\nu)=-\mu\nu/N \mod 1.
$

 The $S$-duality transformation (\ref{SonPsi}) is identical, modulo a modular weight, to the transformation on $\Theta_\mu$ given by (\ref{ThetaStrans}), and therefore, a decomposition such as
\begin{equation}
Z_{\rm ALE}(\tau)
=
\sum_{\mu=0}^{N-1}\Theta_\mu(\tau)
\end{equation}
should not be interpreted too naively as a decomposition of a single scalar partition function into unrelated pieces. Rather, the functions $\Theta_\mu(\tau)$ are naturally the components of a vector-valued boundary state,
\begin{equation}\label{WFALE}
|\Psi(\tau)\rangle
=
\sum_{\mu=0}^{N-1}\Theta_\mu(\tau)|\mu\rangle,
\end{equation}
where
$
\mu\in P/Q\cong \mathbb Z_N
$
labels the flat torsion sector at infinity. Thus, the ALE path integral is more naturally viewed as a vector of sector amplitudes carrying a finite-dimensional representation of the modular group, rather than as a single scalar modular function.

Finally, let us comment on the physical meaning of the indices $\mu$ and $\nu$ appearing in the $\mathbb S$-transformation \eqref{ThetaStrans}. 
Let $c$ denote a generator of the torsion $1$-cycle at the boundary of the ALE space. A flat boundary $U(1)$ connection may then have a nontrivial electric holonomy
\begin{equation}
\operatorname{Hol}_c(a)
=
\exp\left(i\oint_c a\right)
=
\exp\left(\frac{2\pi i\nu}{N}\right),
\qquad
\nu\in\mathbb Z_N.
\end{equation}
Thus, the label $\nu$ may be interpreted as an electric flat-holonomy sector at infinity.
Under electric-magnetic duality, the electric boundary data are exchanged with the corresponding magnetic boundary data, described by the dual gauge field $a_D$. Hence, one may similarly label magnetic flat-holonomy sectors by
\begin{equation}
\operatorname{Hol}_c(a_D)
=
\exp\left(i\oint_c a_D\right)
=
\exp\left(\frac{2\pi i\mu}{N}\right),
\qquad
\mu\in\mathbb Z_N.
\end{equation}

The electric and magnetic torsion variables are not simultaneously diagonalizable. Equivalently, the corresponding Wilson and magnetic operators generate a finite Heisenberg algebra,
\begin{equation}
W_\nu\,H_\mu
=
\exp\left(\frac{2\pi i\mu\nu}{N}\right)
H_\mu\,W_\nu,
\end{equation}
up to a convention-dependent sign in the phase. As a result, a state diagonal in the electric torsion basis is represented in the magnetic basis by a finite Fourier transform.
More explicitly, if $|\nu\rangle$ denotes a basis of electric torsion sectors and $|\mu\rangle_D$ denotes a basis of magnetic torsion sectors, their overlap is
\begin{equation}
{}_D\langle \mu|\nu\rangle
=
\frac{1}{\sqrt N}
\exp\left(-\frac{2\pi i\mu\nu}{N}\right),
\end{equation}
where the sign follows the convention used in \eqref{ThetaStrans}. Therefore, the $S$-duality transformation mixes the sector amplitudes according to
\begin{equation}
\Theta_\mu\!\left(-\frac{1}{\tau}\right)
\sim
\frac{1}{\sqrt N}
\sum_{\nu=0}^{N-1}
\exp\left(-\frac{2\pi i\mu\nu}{N}\right)
\Theta_\nu(\tau),
\end{equation}
up to the overall modular-weight factors discussed above. This is the finite-dimensional analog of the familiar relation in quantum mechanics between position and momentum bases,
$
|p\rangle
=
\int dq\,e^{iqp}|q\rangle.
$
Thus, the indices in \eqref{ThetaStrans} label mutually dual electric and magnetic torsion sectors, and the $\mathbb {S} $-matrix is the discrete Fourier transform between these two polarizations.

\subsection{Gluing Eguchi-Hanson spaces and modularity}
\label{Gluing ALE spaces and modularity}

To test the interpretation of the ALE path integral as preparing the boundary wavefunctional $|\Psi(\tau)\rangle$ in \eqref{WFALE}, one may glue an ALE space $X$ (self-dual in our definition) to its orientation reversal $\bar X$ (anti-self-dual) along their common asymptotic lens-space boundary. In this picture, the path integral on $X$ prepares a state in the Hilbert space associated with the boundary, while the path integral on $\bar X$ prepares the corresponding state that lives in the dual Hilbert space. The path integral on the closed $4$-manifold obtained by gluing,
\begin{equation}
X\cup_{\partial X}\bar X,
\end{equation}
is then given by the inner product
\begin{equation}
\langle \Psi(\bar\tau)|\Psi(\tau)\rangle ,
\end{equation}
and should reproduce the partition function of the theory on the glued $4$-manifold, up to possible normalization factors.

\begin{figure}[t]
    \centering
 \includegraphics[height=0.6\textheight,angle=-90]{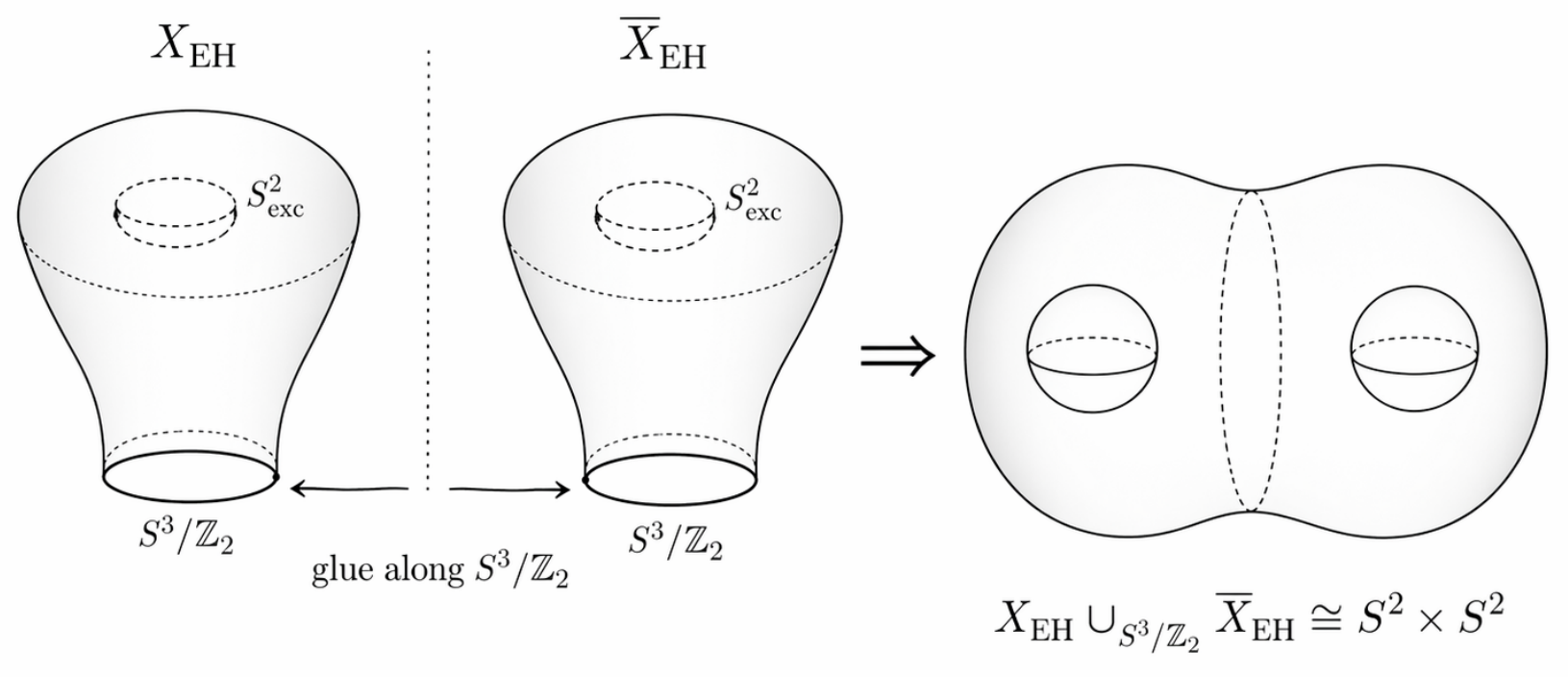}
    \caption{Gluing the Eguchi--Hanson space $X_{\rm EH}$ to its orientation reversal $\overline X_{\rm EH}$ along the common lens-space boundary $S^3/\mathbb Z_2$. The resulting closed $4$-manifold is diffeomorphic to $S^2\times S^2$.}
    \label{fig:EH-gluing}
\end{figure}

We shall test this interpretation in the simplest ALE example, namely the $A_1$ case. The corresponding ALE space is the Eguchi--Hanson space $X_{\rm EH}$, whose asymptotic boundary is the lens space $S^3/\mathbb Z_2$. The boundary sectors are therefore classified by
$
P/Q\cong \mathbb Z_2,
$
which we label by $\mu=0,1$, corresponding to the even and odd sectors. These labels coincide with the possible flat $U(1)$ holonomy classes on the asymptotic lens-space boundary $H^1(S^3/\mathbb Z_2, U(1))=\mathbb Z_2$; see Appendix \ref{app:EH-cohomology-backgrounds}.  Gluing $X_{\rm EH}$ to its orientation reversal $\overline X_{\rm EH}$ along their common $S^3/\mathbb Z_2$ boundary gives a closed $4$-manifold diffeomorphic\footnote{Topologically, the Eguchi--Hanson space is the disk bundle over $S^2$ with Euler number $2$. Its boundary is the lens space $S^3/\mathbb Z_2$. Gluing Eguchi--Hanson space to its oppositely oriented copy along this common boundary gives the double of this disk bundle. The resulting closed $4$-manifold has the same topology as $S^2\times S^2$. In particular, its Betti numbers are
$
b_0=1, b_1=0, b_2=2, b_3=0, b_4=1,
$
so
$
\chi=4,
\sigma=0,
$
in agreement with $S^2\times S^2$. Thus the gluing of the two Eguchi--Hanson blocks produces, topologically, a closed manifold diffeomorphic to $S^2\times S^2$.} to $S^2\times S^2$ \cite{Anderson:1990sgi}; see Figure \ref{fig:EH-gluing}. Thus, if the ALE path integral indeed prepares a boundary wavefunctional, one expects
\begin{equation}
\langle \Psi(\bar\tau)|\Psi(\tau)\rangle
\;\sim\;
Z_{S^2\times S^2}(\tau,\bar\tau).
\end{equation}

From \eqref{WFALE}, the $A_1$ wavefunctional decomposes into two boundary sectors. The gluing pairs a sector on $X_{\rm EH}$ with the conjugate sector on $\overline X_{\rm EH}$. Since for $\mathbb Z_2$ one has $-\mu=\mu$, this gives
\begin{equation}
{}_{\overline X_{\rm EH}}\langle \nu|\mu\rangle_{X_{\rm EH}}
=
\delta_{\mu+\nu,0\;{\rm mod}\;2}
=
\delta_{\mu,\nu}.
\end{equation}
Equivalently, the holonomies on the two sides must match across the gluing. Therefore\footnote{The use of complex conjugation for the theta functions on the orientation-reversed Eguchi--Hanson space may be understood as follows. Reversing the orientation leaves the kinetic term unchanged, but it flips the sign of the topological term, since the intersection form changes sign under orientation reversal (the self-dual space becomes anti-self-dual). Together, these two facts imply that the corresponding theta functions are naturally replaced by their complex conjugates. A more careful derivation of this statement, including the effect of $1$-form symmetry backgrounds, will be given in Section \ref{Gluing ALE spaces with 1-form backgrounds}.}
\begin{equation}\label{EH_inner_product}
\langle \Psi(\bar\tau)|\Psi(\tau)\rangle
=
\overline{\Theta}_0(\bar\tau)\Theta_0(\tau)
+
\overline{\Theta}_1(\bar\tau)\Theta_1(\tau).
\end{equation}
Thus, even sectors are paired with even sectors and odd sectors with odd sectors.

It is convenient to write the two $A_1$ theta functions as sums over integers of fixed parity (these expressions are obtained directly from (\ref{Thetaimport}), recalling that the $A_1$ algebra has a single weight $\bm w_1^2=\frac{1}{2}$):
\begin{equation}
\Theta_\mu(\tau)
=
\sum_{\substack{k\in\mathbb Z\\ k\equiv \mu\;{\rm mod}\;2}}
\exp\left[
\frac{i\pi\tau}{2}k^2
\right],
\qquad
\overline{\Theta}_\mu(\bar\tau)
=
\sum_{\substack{l\in\mathbb Z\\ l\equiv \mu\;{\rm mod}\;2}}
\exp\left[
-\frac{i\pi\bar\tau}{2}l^2
\right].
\end{equation}
Using
$
\tau=\frac{\theta}{2\pi}+i\frac{4\pi}{g^2},\,
\bar\tau=\frac{\theta}{2\pi}-i\frac{4\pi}{g^2},
$
the exponent in the product becomes
$
\frac{i\pi\tau}{2}k^2
-
\frac{i\pi\bar\tau}{2}l^2
=
-\frac{2\pi^2}{g^2}(k^2+l^2)
+
\frac{i\theta}{4}(k^2-l^2).
$
In \eqref{EH_inner_product}, the two terms together sum over pairs $(k,l)$ with the same parity. We may therefore introduce integer variables
$
p=\frac{k-l}{2},\,
q=\frac{k+l}{2}.
$
This change of variables is a bijection between pairs $(k,l)$ of the same parity and pairs $(p,q)\in\mathbb Z^2$. In terms of $p$ and $q$ we have
$
k^2+l^2=2(p^2+q^2),\,
k^2-l^2=4pq.
$
Hence
\begin{eqnarray}\nonumber\langle \Psi(\bar\tau)|\Psi(\tau)\rangle&=&Z_{\rm glue}(\tau,\bar\tau)
=
\overline{\Theta}_0(\bar\tau)\Theta_0(\tau)
+
\overline{\Theta}_1(\bar\tau)\Theta_1(\tau)\\
&=&
\sum_{p,q\in\mathbb Z}
\exp\left[
-\frac{4\pi^2}{g^2}(p^2+q^2)
+
i\theta pq
\right].
\end{eqnarray}
This is precisely the flux-sum form of the Maxwell classical partition function on $S^2\times S^2$ given in (\ref{ZS2S2}).
Thus, modulo the overall normalization of the path integral, the inner product of the wavefunctional prepared by Eguchi--Hanson space with its conjugate reproduces the partition function on the glued closed manifold $S^2\times S^2$.\footnote{The gluing of two Eguchi--Hanson spaces fixes the topology of the resulting manifold, but not its metric. The equal-radii partition function on $S^2\times S^2$ arises for the simplest symmetric gluing, in which the two ALE pieces are equipped with identical metrics and are glued with a symmetric identification along the common lens-space boundary.}

Under $S$-duality, the two $A_1$ theta functions transform as a vector according to (\ref{ThetaStrans}):
\begin{equation}
\begin{pmatrix}
\Theta_0(-1/\tau)\\[2mm]
\Theta_1(-1/\tau)
\end{pmatrix}
=
\frac{(-i\tau)^{1/2}}{\sqrt2}
\begin{pmatrix}
1&1\\
1&-1
\end{pmatrix}
\begin{pmatrix}
\Theta_0(\tau)\\[2mm]
\Theta_1(\tau)
\end{pmatrix}.
\end{equation}
The antiholomorphic theta functions transform by the complex-conjugate rule,
\begin{equation}
\begin{pmatrix}
\overline{\Theta}_0(-1/\bar\tau)\\[2mm]
\overline{\Theta}_1(-1/\bar\tau)
\end{pmatrix}
=
\frac{(i\bar\tau)^{1/2}}{\sqrt2}
\begin{pmatrix}
1&1\\
1&-1
\end{pmatrix}
\begin{pmatrix}
\overline{\Theta}_0(\bar\tau)\\[2mm]
\overline{\Theta}_1(\bar\tau)
\end{pmatrix}.
\end{equation}
Using the unitarity of the finite Fourier matrix, the glued partition sum transforms as
\begin{equation}
\begin{aligned}
Z_{\rm glue}(-1/\tau,-1/\bar\tau)
&=
\overline{\Theta}_0(-1/\bar\tau)\Theta_0(-1/\tau)
+
\overline{\Theta}_1(-1/\bar\tau)\Theta_1(-1/\tau)
\\
&=
|\tau|\,
\left[
\overline{\Theta}_0(\bar\tau)\Theta_0(\tau)
+
\overline{\Theta}_1(\bar\tau)\Theta_1(\tau)
\right].
\end{aligned}
\end{equation}
Thus, the glued theta sum transforms as
\begin{equation}
Z_{\rm glue}(-1/\tau,-1/\bar\tau)
=
|\tau|\,Z_{\rm glue}(\tau,\bar\tau).
\end{equation}
This is precisely the expected modular behavior of the partition function on $S^2\times S^2$ as in (\ref{S2S2duality}). 

Under $\mathbb T$, the $A_1$ theta functions transform diagonally according to (\ref{Tdualitytrans}):
\begin{equation}
\Theta_0(\tau+1)=\Theta_0(\tau),
\qquad
\Theta_1(\tau+1)=i\,\Theta_1(\tau).
\end{equation}
The antiholomorphic theta functions transform with the complex-conjugate phases,
\begin{equation}
\overline{\Theta}_0(\bar\tau+1)=\overline{\Theta}_0(\bar\tau),
\qquad
\overline{\Theta}_1(\bar\tau+1)=-i\,\overline{\Theta}_1(\bar\tau),
\end{equation}
Consequently, the  glued theta sum is invariant under $\mathbb T$-transformation,
\begin{equation}
Z_{\rm glue}(\tau+1,\bar\tau+1)
=
Z_{\rm glue}(\tau,\bar\tau)
\end{equation}
as required on a closed spin manifold.

A possible concern is that, up to this point, we have compared the partition function obtained by gluing the two Eguchi--Hanson spaces only with the classical flux sum on $S^2\times S^2$, rather than with the full quantum partition function. Let us now address this point. For free Maxwell theory on a compact closed manifold, the Gaussian integral over fluctuations contributes an overall factor $g^{\,b_0-b_1}$, where $b_0$ and $b_1$ are the Betti numbers of the manifold. In the case of $S^2\times S^2$, one has $b_0=1$ and $b_1=0$, so the $1$-loop normalization contributes a single power of $g$. By contrast, on an ALE space such as Eguchi--Hanson, the relevant $1$-loop determinant is defined using normalizable modes, and there are no normalizable zero modes of either the scalar or $1$-form Laplacians. One may therefore adopt a regularization scheme on each ALE block for which no intrinsic power of $g$ remains. From this viewpoint, the single factor of $g$ needed on the glued manifold should be attributed to the normalization associated with gluing across the common boundary, rather than to bulk zero modes of the individual Eguchi--Hanson pieces. It is precisely this additional factor that restores the full modular weight of the $S^2\times S^2$ partition function, as in (\ref{S2S2dualityfull}).

Can this gluing construction be generalized to any ALE space? A natural conjecture is that gluing an $A_{N-1}$ ALE space to its orientation reversal yields a closed $4$-manifold with intersection form equivalent to that of the connected sum $\#^{\,N-1}(S^2\times S^2)$. Whether the resulting smooth manifold is literally diffeomorphic to $\#^{\,N-1}(S^2\times S^2)$ is a more refined question beyond the scope of this work.

\section{Electric and magnetic $1$-form symmetry backgrounds and their gauging}
\label {Electric and magnetic 1-form symmetry backgrounds and their gauging}

We now introduce background fields for the electric and magnetic $1$-form symmetries. In the formulation in terms of the $U(1)$ gauge field $A$, a background for the electric $1$-form symmetry is incorporated through the gauge-invariant combination
\begin{equation}
F \;\longrightarrow\; F-B_e,
\qquad F=dA,
\end{equation}
where $B_e$ is a background $2$-form connection. The fields $A$ and $B_e$ transform under a $U(1)^{[1]}$ gauge transformation as
\begin{equation}
A\;\longrightarrow\; A+\Lambda,
\qquad
B_e\;\longrightarrow\; B_e+d\Lambda,
\end{equation}
so that $F-B_e$ is invariant.

Likewise, a background for the magnetic $1$-form symmetry may be introduced through a linear coupling to a second background $2$-form connection $B_m$. In the presence of both backgrounds, the classical action is naturally deformed to
\begin{equation}\label{bckaction}
S_\tau
=
\frac{1}{g^{2}}\int (F-B_e)\wedge \star (F-B_e)
-
\frac{i\theta}{8\pi^{2}}\int (F-B_e)\wedge (F-B_e)
-
\frac{i}{2\pi}\int B_m\wedge (F-B_e)\,.
\end{equation}
In this form, $B_e$ shifts the flux lattice, while $B_m$ couples linearly to the fluxes.

The continuous $U(1)$ $1$-form symmetry backgrounds $B_e$ and $B_m$ are $U(1)$ $2$-form connections whose essential data are carried by their periods through the compact bulk $2$-cycles. From the universal coefficient theorem, we have
 $H^2(X,U(1))
\cong U(1)^{N-1}
$; see Appendix \ref{app:EH-cohomology-backgrounds} for a detailed discussion of the Eguchi-Hanson, $A_1$ algebra, case.  Thus, the continuous backgrounds are naturally parametrized by $U(1)$-valued periods on the compact exceptional cycles. By contrast, there is no additional independent continuous holonomy at asymptotic infinity, since the asymptotic boundary $S^3/\mathbb Z_N$ has no nontrivial ordinary $2$-cycles, $H_2(S^3/\mathbb Z_N,\mathbb Z)=0$, implying $H^2(S^3/\mathbb Z_N,U(1))=0$. When one restricts the continuous $U(1)$ $1$-form symmetry to a discrete subgroup $\mathbb Z_k$, the background becomes a discrete $\mathbb Z_k$-valued $2$-form gauge field, and the bulk sectors are correspondingly labeled by classes in $H^2(X,\mathbb Z_k)\cong (\mathbb Z_k)^{N-1}$. These discrete data again measure the fluxes through the compact bulk $2$-cycles. Their restriction to the asymptotic lens-space boundary does not produce a new continuous holonomy, but only a torsion class in
$
H^2(S^3/\mathbb Z_N,\mathbb Z_k)\cong \mathbb Z_{\gcd(N,k)}.
$
Hence, the continuous $1$-form symmetry backgrounds are controlled by bulk periods, while after restriction to $\mathbb Z_k$ they become discrete bulk flux sectors together with, at most, a residual boundary torsion determined by the greatest common divisor of $N$ and $k$.

To make the discussion explicit, recall that $\{\Sigma_i\}$, $i=1,\dots,N-1$ are the $2$-cycles of the ALE space and that $\{F_i\}$ are basis of closed $2$-forms dual to them, normalized by
\begin{equation}
\int_{\Sigma_j}F_i=2\pi\delta_{ij}.
\end{equation}
Thus, a Maxwell flux and the continuous electric and magnetic $1$-form symmetry backgrounds may be expanded as
\begin{equation}
F=\sum_{i=1}^{N-1}m_iF_i,
\qquad
B_e=\sum_{i=1}^{N-1}b_iF_i,
\qquad
B_m=\sum_{i=1}^{N-1}c_iF_i,
\end{equation}
and the sector functions are therefore generalized to
\begin{eqnarray}\nonumber
\Theta_\mu(\tau;b,c)
=
\sum_{\substack{m_i\in\mathbb Z\\ \sum_{j=1}^{N-1} j\,m_j\equiv \mu\;(\mathrm{mod}\,N)}}
\exp\!\left[
i\pi\tau
\sum_{i,j=1}^{N-1}
(m_i-b_i)(C^{-1})_{ij}(m_j-b_j)
\right.\\
\left.+
2\pi i
\sum_{i,j=1}^{N-1}
c_i(C^{-1})_{ij}(m_j-b_j)
\right].
\end{eqnarray}
Here, the first term arises from the quadratic Maxwell action with the electric background $B_e$, while the second term is induced by the linear coupling to the magnetic background $B_m$.
Equivalently, in lattice notation, one may write
\begin{equation}\label{thetabackground}
\Theta_\mu(\tau;b,c)
=
\sum_{\lambda\in \mathcal C_\mu}
\exp\!\left(
i\pi\tau\,(\lambda-b)^2
+
2\pi i\,c\!\cdot\!(\lambda-b)
\right),
\end{equation}
where $\mathcal C_\mu$ denotes the coset of the root lattice corresponding to the sector of $N$-ality $\mu$. More explicitly, one may parametrize
\begin{equation}
\lambda
=
\lambda_\mu+\sum_{a=1}^{N-1}m_a\,\bm\alpha_a,
\end{equation}
with $\lambda_0=0$ and $\lambda_\mu=\bm w_\mu$ for $\mu=1,\dots,N-1$. In this form, the role of the two backgrounds is transparent: the electric background $b$ shifts the lattice, while the magnetic background $c$ couples linearly to the flux.

The continuous electric and magnetic $1$-form symmetry backgrounds are naturally valued in the real Cartan torus
$
\mathfrak h_{\mathbb R}/Q\simeq U(1)^{N-1},
$
where $Q$ is the root lattice of $A_{N-1}$. Thus, if
$
b=\sum_{i=1}^{N-1}b_i\bm w_i,\,
 c=\sum_{i=1}^{N-1}c_i\bm w_i,
$
then large background gauge transformations identify
\begin{equation}
 b\sim  b+\bm\alpha,
\qquad
 c\sim c+\bm\alpha',
\qquad
\bm\alpha,\bm\alpha'\in Q.
\end{equation}
These identifications specify the periodicities of the background parameters. In the Eguchi--Hanson case, corresponding to $A_1$, the root lattice is generated by
$
\bm\alpha_1=2\bm w_1,
$
and one obtains
\begin{equation}
b\sim b+2,
\qquad
c\sim c+2,
\end{equation}
so that $b,c\in \mathbb R/2\mathbb Z$ in this normalization.

There is, however, an important distinction between the two backgrounds. The electric background is strictly periodic under root-lattice shifts,
\begin{equation}
\Theta_\mu(\tau; b+\bm\alpha, c)
=
\Theta_\mu(\tau; b, c),
\qquad
\bm\alpha\in Q,
\end{equation}
because such a shift can be absorbed by a relabeling of the flux lattice. By contrast, the magnetic background is only quasi-periodic:
\begin{equation}
\Theta_\mu(\tau; b, c+\bm\alpha)
=
e^{-2\pi i\,\bm\alpha\cdot  b}\,
\Theta_\mu(\tau; b, c),
\qquad
\bm\alpha\in Q,
\end{equation}
with the sign convention of Eq. (\ref{thetabackground}). This quasi-periodicity is the background-field manifestation of the mixed electric-magnetic $1$-form anomaly. Equivalently, the ALE theta block is not an ordinary function on the background Cartan torus, but a section of a line bundle over the space of $( b, c)$ backgrounds.

With both backgrounds turned on, the family of functions $\Theta_\mu(\tau;b,c)$ is closed under modular transformations. Under $\mathbb T:\tau\mapsto\tau+1$, one finds
\begin{equation}\label{TtransBack}
\mathbb T:\quad
\Theta_\mu(\tau;b,c)\mapsto \Theta_\mu(\tau+1;b,c)
=
e^{\,i\pi\mu^2\left(1-\frac1N\right)}
e^{-\,i\pi b^2}\,
\Theta_\mu(\tau;b,c-b).
\end{equation}
Thus $\mathbb T$ acts diagonally on the sector label $\mu$, while shifting the magnetic background variable by the electric one.
Under $\mathbb S:\tau\mapsto -1/\tau$, Poisson resummation yields\footnote{The modular transformations should be understood together with the large background gauge transformations.  Indeed, the phase generated by a root-lattice shift of one background is canceled by the quasi-periodicity of the transformed theta block.}
\begin{equation}\label{SBACK}
\mathbb S:\quad
\Theta_\mu(\tau;b,c)\mapsto
\Theta_\mu\!\left(-\frac1\tau;b,c\right)
=
\frac{(-i\tau)^{\frac{N-1}{2}}}{\sqrt N}\,
e^{-\,2\pi i\,b\cdot c}
\sum_{\nu=0}^{N-1}
e^{-\,\frac{2\pi i\mu\nu}{N}}\,
\Theta_\nu(\tau;c,-b).
\end{equation}
Hence, the $\mathbb {S} $-transformation exchanges the electric and magnetic background data, up to the expected sign, while acting on the sector label by the discrete Fourier transform\footnote{It is useful to distinguish the action of the modular group on charges from its induced action on background sources. Under $\mathbb T$, the electric-magnetic charge vector transforms as
$
(q_e,q_m)\mapsto(q_e+q_m,q_m),
$
whereas the background fields transform in the dual way
$
(b,c)\mapsto(b,c-b).
$
Similarly, under $\mathbb S$ the charges transform as
$
(q_e,q_m)\mapsto(q_m,-q_e),
$
while the corresponding backgrounds transform as
$
(b,c)\mapsto(c,-b).
$
Thus, the background fields transform in the dual representation to the charge lattice, as expected for sources coupled to the corresponding $1$-form symmetry currents.}. By applying $\mathbb S$ and $\mathbb T$ iteratively, one can also show 
\begin{eqnarray}
(\mathbb S\mathbb T^{-1})^3\Theta_\mu(\tau;b,c)=\mathbb S^2\Theta_\mu(\tau;b,c)=\Theta_{-\mu}(\tau;-b,-c)\,.
\end{eqnarray}
Again, this implies that $\mathbb S^2$ acts as a charge conjugation operation by negating the $N$-alities of the sectors as well as the electric and magnetic backgrounds. 

As we pointed out above, there are two special fixed points in the upper-half complex plane at $\tau_\star=i$ and $\tau_{\star\star}=e^{i\pi/3}$, under the action of $\mathbb S$ and $\mathbb S\mathbb T^{-1}$, respectively. At these points, the action of $\mathbb S$ and $\mathbb S\mathbb T^{-1}$ becomes actual symmetries, $\mathbb Z_4^{(0)}$ and $\mathbb Z_6^{(0)}$, respectively, and it is interesting to understand how these symmetries act in the presence of the electric and magnetic backgrounds.

 At $\tau_\star$, and in the presence of $1$-form symmetry backgrounds, $\mathbb S\equiv \mathbb Z_4^{(0)}$ acts by
\begin{equation}
 \mathbb Z_4^{(0)}:\quad\Theta_\mu\!\left(\tau_\star;b,c\right)
\;\mapsto\;
\frac{1}{\sqrt N}\,
e^{-2\pi i\,b\cdot c}
\sum_{\nu=0}^{N-1}
e^{-\,\frac{2\pi i\mu\nu}{N}}\,
\Theta_\nu(\tau_\star;c,-b).
\end{equation}
Its action is not diagonal in the electric torsion basis. Instead,  it mixes the sector amplitudes by a discrete Fourier transform while rotating the electric and magnetic backgrounds as
$
(b,c)\mapsto(c,-b).
$
The mixing of sectors is part of the ordinary symmetry action on the vector of sector amplitudes, whereas the background-dependent factor $e^{-2\pi i\,b\cdot c}$ reflects the mixed anomaly between the electric and magnetic $1$-form symmetries.

At the second fixed point $\tau_{\star\star}=e^{i\pi/3}$, which satisfies $-1/(\tau_{\star\star}-1)=\tau_{\star\star}$, the generator $\mathbb S\mathbb T^{-1}\equiv \mathbb Z_6^{(0)}$ acts on the background-coupled sector functions as: 
\begin{equation}
\begin{split}
\Theta_\mu\!\left(\tau_{\star\star};b,c\right)
=\;&
\frac{\left[-i\left(\tau_{\star\star}-1\right)\right]^{\frac{N-1}{2}}}{\sqrt N}\,
e^{-2\pi i\,b\cdot c}\,
e^{+i\pi c^2}
\\
&\times
\sum_{\nu=0}^{N-1}
e^{-\,\frac{2\pi i\mu\nu}{N}}\,
e^{-\,i\pi\nu^2\left(1-\frac1N\right)}\,
\Theta_\nu\!\left(\tau_{\star\star};\,c,\,c-b\right),
\end{split}
\label{Z6fixedpointbackground}
\end{equation}
where $\left[-i(\tau_{\star\star}-1)\right]^{\frac{N-1}{2}}=e^{i\pi(N-1)/12}$. 
It is instructive to rewrite (\ref{Z6fixedpointbackground}), with the transformed background data on the left-hand side. Relabeling the backgrounds gives
\begin{equation}
\Theta_\mu\!\left(\tau_{\star\star};\,b-c,\,b\right)
=
\frac{e^{i\pi(N-1)/12}}{\sqrt N}\,
e^{-i\pi b^2}\,e^{+2\pi i\,b\cdot c}
\sum_{\nu=0}^{N-1}
e^{-\,\frac{2\pi i\mu\nu}{N}}\,
e^{-\,i\pi\nu^2\left(1-\frac1N\right)}\,
\Theta_\nu\!\left(\tau_{\star\star};\,b,\,c\right).
\label{Z6active}
\end{equation}
Thus, $\mathbb Z_6^{(0)}$ acts on the sector labels by the discrete Fourier transform dressed with the $\mathbb T^{-1}$ phases, and on the $1$-form backgrounds by
\begin{equation}
(b,c)\longmapsto(b-c,\,b).
\end{equation}

The Fourier matrix is the ordinary linear action of the duality symmetry on the boundary-state vector $\{\Theta_\mu\}$. The factor $e^{-i\pi b^2}e^{+2\pi i\,b\cdot c}$ is induced by the $1$-form symmetry backgrounds. An important feature of (\ref{Z6active}) is that $\mathbb Z_6^{(0)}$ does not preserve the purely electric backgrounds. Setting $c=0$, one finds
\begin{equation}
\Theta_\mu\!\left(\tau_{\star\star};\,b,\,b\right)
=
\frac{e^{i\pi(N-1)/12}}{\sqrt N}\,e^{-i\pi b^2}
\sum_{\nu=0}^{N-1}
e^{-\,\frac{2\pi i\mu\nu}{N}}\,
e^{-\,i\pi\nu^2\left(1-\frac1N\right)}\,
\Theta_\nu\!\left(\tau_{\star\star};\,b,\,0\right),
\end{equation}
so a purely electric background is mapped to a dyonic one, $(b,0)\mapsto(b,b)$, accompanied by the quadratic phase $e^{-i\pi b^2}$.  Consequently, unless the electrically and dyonically gauged theories are related by an additional identification or an admissible local counterterm, $\mathbb Z_6^{(0)}$ does not survive as an ordinary invertible symmetry of the electrically gauged theory. We leave a systematic treatment to future work.

\subsection{Gluing Eguchi--Hanson spaces with $1$-form backgrounds}
\label{Gluing ALE spaces with 1-form backgrounds}

We now refine the gluing discussion of Section \ref{Gluing ALE spaces and modularity}  by allowing independent electric and magnetic $1$-form backgrounds on the two Eguchi--Hanson pieces before gluing. This will allow us to recover the most general background-dependent partition function on the closed manifold $S^2\times S^2$, providing more evidence for the interpretation of $\Theta_\mu$ as the amplitudes of boundary sectors $\{|\mu\rangle\}$.

We begin with a comment about orientation reversal, which we slipped under the rug in Section \ref{Gluing ALE spaces and modularity}, which now needs to be addressed. On the Eguchi--Hanson space $X_{\rm EH}$, the relevant lattice pairing is determined by the inverse Cartan matrix of $A_1$, which in our normalization is simply
$
\bm w_1^2=\frac12.
$
Upon reversing the orientation, the self-dual and anti-self-dual spaces are exchanged, and correspondingly, the intersection form changes sign. Equivalently, the inverse Cartan pairing appearing in the theta function is reversed:
$
(\cdot,\cdot)_{\overline X_{\rm EH}}=-(\cdot,\cdot)_{X_{\rm EH}}.
$
Thus, if we keep the background labels themselves fixed across the gluing, then the effect of orientation reversal is implemented by changing the sign of the quadratic and bilinear forms that enter the theta functions on $\overline X_{\rm EH}$. 

For the $A_1$ ALE space, the boundary Hilbert space is
$
\mathcal H_{\partial X_{\rm EH}}
=
\mathrm{span}\{|0\rangle,|1\rangle\},
$
and the corresponding lattice cosets are
$
\mathcal C_\mu=\{(2m+\mu)\bm w_1\mid m\in\mathbb Z\},
\mu\in\mathbb Z_2.
$
We now allow independent electric ($b$) and magnetic ($c$) background data on the right and left Eguchi--Hanson pieces:
\begin{eqnarray}\nonumber
b_R&=&x_R\,\bm w_1,
\qquad
c_R=y_R\,\bm w_1,\quad \mbox{on} \quad X_{\rm EH}\,,\\
b_L&=&x_L\,\bm w_1,
\qquad
c_L=y_L\,\bm w_1, \quad \mbox{on} \quad \overline X_{\rm EH}\,.
\end{eqnarray}

On the right piece, the sector functions are
\begin{align}\label{RightThetaBCK}
\Theta^{(R)}_\mu(\tau;x_R,y_R)
&=
\sum_{\lambda\in\mathcal C_\mu}
\exp\!\left(
i\pi\tau\,(\lambda-b_R)^2
+
2\pi i\,c_R\!\cdot\!(\lambda-b_R)
\right)
\nonumber\\
&=
\sum_{n\in\mathbb Z}
\exp\!\left[
\frac{i\pi\tau}{2}(2n+\mu-x_R)^2
+
i\pi y_R(2n+\mu-x_R)
\right].
\end{align}
On the orientation-reversed piece, the same background labels are used, but the pairing is reversed. Thus
\begin{align}
\Theta^{(L)}_\mu(\bar\tau;x_L,y_L)
&=
\sum_{\lambda\in\mathcal C_\mu}
\exp\!\left(
i\pi\bar\tau\,(\lambda-b_L)^2_{\overline X}
+
2\pi i\,c_L\!\cdot_{\overline X}\!(\lambda-b_L)
\right)
\nonumber\\
&=
\sum_{m\in\mathbb Z}
\exp\!\left[
-\frac{i\pi\bar\tau}{2}(2m+\mu-x_L)^2
-
i\pi y_L(2m+\mu-x_L)
\right].
\end{align}
Now, we see that $\Theta^{(L)}_\mu(\bar\tau;x_L,y_L)$ is nothing but $\Theta^{(R)}_\mu(\tau;x_L,y_L)$ after complex conjugating and replacing right with left backgrounds, and thus, we recover the same structure of the theta functions we constructed in Section  \ref{Gluing ALE spaces and modularity} after switching off the backgrounds.
The wavefunctionals prepared by the two pieces are therefore
\begin{eqnarray}\nonumber
|\Psi_R(\tau;x_R,y_R)\rangle
=
\sum_{\mu\in\mathbb Z_2}
\Theta^{(R)}_\mu(\tau;x_R,y_R)\,|\mu\rangle,\quad
\langle \Psi_L(\bar\tau;x_L,y_L)|
=
\sum_{\nu\in\mathbb Z_2}
\,\Theta^{(L)}_\nu(\bar\tau;x_L,y_L)\langle \nu|\,.\\
\end{eqnarray}

Upon gluing, one must distinguish two kinds of boundary data. The first is the ordinary flat holonomy of the dynamical $U(1)$ gauge field at the lens-space boundary. Since the field strength becomes flat asymptotically, the boundary value of the gauge field is classified by
$
H^1(S^3/\mathbb Z_2,U(1))
\simeq
\mathrm{Hom}(H_1(S^3/\mathbb Z_2,\mathbb Z),U(1))
\simeq
\mathbb Z_2 .
$
This is precisely the boundary-sector label $\mu$ appearing in the Eguchi--Hanson theta blocks, and gluing requires the usual pairing of equal holonomy sectors,
$
\langle \mu|\nu\rangle=\delta_{\mu\nu}.
$
The second kind of data comes from the background fields for the electric and magnetic $1$-form symmetries. For continuous $U(1)$ $1$-form symmetry backgrounds, these are $U(1)$ $2$-form connections. Their possible boundary classes would lie in
$
H^2(S^3/\mathbb Z_2,U(1)).
$
However,
$
H^2(S^3/\mathbb Z_2,U(1))=0,
$
so the continuous $1$-form symmetry backgrounds carry no additional independent boundary holonomy on the lens space; see Appendix \ref{app:EH-cohomology-backgrounds} for a detailed explanation. Therefore, in the presence of continuous backgrounds $B_e$ and $B_m$, the gluing proceeds as in the case without such backgrounds: one matches the dynamical $U(1)$ holonomy sectors $\mu$, while the background periods are related across the two Eguchi--Hanson pieces by the same bulk-to-glued-manifold change of variables.

With the diagonal gluing pairing
$
\langle \nu|\mu\rangle=\delta_{\nu,\mu},
$
the glued partition function is
\begin{align}
Z_{\rm glue}(\tau,\bar\tau;x_R,x_L;y_R,y_L)
&=
\langle \Psi_L(\bar\tau;x_L,y_L)\mid \Psi_R(\tau;x_R,y_R)\rangle
\nonumber\\
&=
\sum_{\mu=0}^1
\Theta^{(L)}_\mu(\bar\tau;x_L,y_L)\,
\Theta^{(R)}_\mu(\tau;x_R,y_R).
\end{align}
Substituting the explicit expressions gives
\begin{align}
Z_{\rm glue}
&=
\sum_{\mu=0}^1
\sum_{m,n\in\mathbb Z}
\exp\!\left[
-\frac{2\pi^2}{g^2}\Big((2m+\mu-x_L)^2+(2n+\mu-x_R)^2\Big)
\right]
\nonumber\\
&\hspace{0.7cm}\times
\exp\!\left[
\frac{i\theta}{4}\Big((2n+\mu-x_R)^2-(2m+\mu-x_L)^2\Big)
\right]
\nonumber\\
&\hspace{0.7cm}\times
\exp\!\left[
i\pi y_R(2n+\mu-x_R)-i\pi y_L(2m+\mu-x_L)
\right].
\end{align}

It is now convenient to introduce
$
k=2n+\mu,
l=2m+\mu.
$
Because $\mu$ is summed over $\{0,1\}$, the pair $(k,l)$ runs over all integers of the same parity. We then define
$
p=\frac{k-l}{2},
q=\frac{k+l}{2}.
$
Since $k$ and $l$ have the same parity, both $p$ and $q$ are integers, and the map $(m,n,\mu)\mapsto (p,q)$ is bijective
\footnote{Also, a short computation gives
$
(l-x_L)^2+(k-x_R)^2
=
2\left(p-\frac{x_R-x_L}{2}\right)^2
+
2\left(q-\frac{x_R+x_L}{2}\right)^2,
$
$
(k-x_R)^2-(l-x_L)^2
=
4\left(p-\frac{x_R-x_L}{2}\right)
\left(q-\frac{x_R+x_L}{2}\right),
$
and
$
y_R(k-x_R)-y_L(l-x_L)
=
(y_R+y_L)\left(p-\frac{x_R-x_L}{2}\right)
+
(y_R-y_L)\left(q-\frac{x_R+x_L}{2}\right).
$}.
Using the identifications
\begin{eqnarray}\label{leftrightident}\nonumber
x_1&=&-\frac{x_R-x_L}{2},
\qquad
x_2=-\frac{x_R+x_L}{2},\\
y_1&=&\frac{y_R+y_L}{2},
\qquad
y_2=\frac{y_R-y_L}{2},
\end{eqnarray}
 and substituting back,  the glued partition function becomes
\begin{align}
Z_{\rm glue}(\tau,\bar\tau;x_1,x_2;y_1,y_2)
&=
\sum_{p,q\in\mathbb Z}
\exp\!\left[
-\frac{4\pi^2}{g^2}\Big((p+x_1)^2+(q+x_2)^2\Big)
\right]
\nonumber\\
&\qquad\times
\exp\!\left[
i\theta\,(p+x_1)(q+x_2)
\right]
\nonumber\\
&\qquad\times
\exp\!\left[
2\pi i\big(y_1(p+x_1)+y_2(q+x_2)\big)
\right].
\label{eq:S2S2-full-background}
\end{align}

This is now to be compared with the general Maxwell partition function on $S^2\times S^2$ in the presence of electric and magnetic $1$-form backgrounds (with the understanding that we ignored the Gaussian fluctuations, which contribute an overall factor of $g$). If $\Sigma_1$ and $\Sigma_2$ denote the two generating $2$-spheres, then a general electric background is specified by its periods 
\begin{equation}
x_1=-\int_{\Sigma_2}\frac{B_e}{2\pi},
\qquad
x_2=-\int_{\Sigma_1}\frac{B_e}{2\pi},
\end{equation}
and similarly a magnetic background by
\begin{equation}
y_1=\int_{\Sigma_1}\frac{B_m}{2\pi},
\qquad
y_2=\int_{\Sigma_2}\frac{B_m}{2\pi}.
\end{equation}
Shifting the electric and magnetic fluxes by these backgrounds in the action of Maxwell theory on $S^2\times S^2$ (\ref{actionS2}), and after adding the coupling between the $1$-form electric and magnetic backgrounds, $\frac{-i}{2\pi}\int B_m\wedge (F-B_e)$, we recover (\ref{eq:S2S2-full-background}).

In this way, the full four-parameter family of $1$-form symmetry backgrounds on $S^2\times S^2$ is naturally recovered from the gluing of the two Eguchi--Hanson halves, provided one allows independent background data on the two sides. 

For the glued partition function on $S^{2}\times S^{2}$ in electric and magnetic $1$-form backgrounds,
the modular generators act by background rotation together with the expected automorphy factors. Under $\mathbb T$, one finds
\begin{equation}
Z_{\rm glue}(\tau+1,\bar\tau+1;x_1,x_2;y_1,y_2)
=
e^{-2\pi i x_1x_2}\,
Z_{\rm glue}(\tau,\bar\tau;x_1,x_2;\,y_1+x_2,\;y_2+x_1),
\end{equation}
and under $\mathbb S$,
\begin{equation}
Z_{\rm glue}\!\left(-\frac1\tau,-\frac1{\bar\tau};x_1,x_2;y_1,y_2\right)
=
|\tau|\,
e^{2\pi i(x_1y_1+x_2y_2)}\,
Z_{\rm glue}(\tau,\bar\tau;\,-y_2,-y_1;\,x_2,x_1).
\end{equation}
These are precisely the same transformation laws obtained by first acting on the two Eguchi--Hanson blocks separately and then gluing them\footnote{Again, as in Section \ref{Gluing ALE spaces and modularity}, in writing $Z_{\rm glue}$ we neglected a multiplicative factor of $g$ resulting from performing the Gaussian integral, which should be restored after gluing.}.

We now turn to gluing in the presence of discrete $\mathbb Z_k$ $1$-form symmetry backgrounds. In this case, there is a subtlety that is absent for continuous $U(1)$ backgrounds. A continuous $U(1)$ $2$-form background has no independent boundary class on the Eguchi--Hanson lens-space boundary, since
$
H^2(S^3/\mathbb Z_2,U(1))=0.
$
By contrast, a discrete $\mathbb Z_k$ $2$-form background can have a torsion boundary restriction. For $Y=S^3/\mathbb Z_2$, one has
$
H^2(Y,\mathbb Z_k)
\simeq
\mathbb Z_{\gcd(2,k)} ,
$
as reviewed in Appendix \ref{app:EH-cohomology-backgrounds}.
This immediately separates the odd- and even-$k$ cases. If $k$ is odd, then
$
H^2(Y,\mathbb Z_k)=0,
$
so the discrete background carries no extra boundary datum. The gluing is then directly analogous to the continuous-background case: one matches the dynamical $U(1)$ flat holonomy sectors $\mu$, and the two bulk $\mathbb Z_k$ labels on the Eguchi--Hanson pieces give a complete parametrization of the closed-manifold backgrounds in
$
H^2(S^2\times S^2,\mathbb Z_k)\simeq \mathbb Z_k^2.
$
For even $k$, however,
\begin{equation}
H^2(Y,\mathbb Z_k)\simeq \mathbb Z_2 .
\end{equation}
Thus, a $\mathbb Z_k$ background on an Eguchi--Hanson half has an additional boundary parity. If the bulk labels on the two halves are
$
r_L,r_R\in\mathbb Z_k,
$
their restrictions to the common boundary are
$
r_L \!\!\mod 2
$
and
$
r_R \!\!\mod 2.
$
Ordinary gluing requires these boundary parities to be compatible. When $k$ is even, the gluing map from the two absolute Eguchi--Hanson labels to the closed-manifold background labels does not cover all of
$
H^2(S^2\times S^2,\mathbb Z_k).
$
Then, if one gauges $\mathbb Z_k$ (meaning summing over $\mathbb Z_k$-valued $2$-form connections as we do in the next section), the naive gluing of gauged ALE blocks is obstructed/not sufficient to reconstruct the full closed-manifold gauging.

\subsection{Gauging a $\mathbb Z_k$ subgroup of the $1$-form symmetries on Eguchi--Hanson space}

We now discuss the gauging of a discrete $\mathbb Z_k$ subgroup of the electric and magnetic $1$-form symmetries in the Eguchi--Hanson, or $A_1$, case. The gauging is implemented by summing over the corresponding $\mathbb Z_k$-valued $2$-form background sectors, possibly with the insertion of discrete theta-angle phases and Fourier characters. As discussed above, even $k$ is problematic, so we restrict to odd values of $k$. 

For the $A_1$ Eguchi--Hanson block, we use $r$ and $s$ instead of $b$ and $c$, for the electric and magnetic backgrounds, respectively. In the period-$2$ normalization for the background used for Eguchi--Hanson (recall that the electric and magnetic background parameters are periodic modulo the root lattice $Q=2\mathbb Z\,\bm w_1$), this $\mathbb Z_k$ subgroup is represented by
\begin{equation}
b_r=\frac{2r}{k},
\qquad c_r=\frac{2s}{k}\,,\quad
r,s\in\mathbb Z_k.
\end{equation}
and the theta functions are
\begin{equation}
\Theta_\mu(\tau;r,s)
:=
\Theta_\mu\!\left(\tau;\frac{2r}{k},\frac{2s}{k}\right),
\qquad
\mu=0,1,
\qquad
r,s\in \mathbb Z_k,
\end{equation}
and the corresponding ALE state is
\begin{equation}
|\Psi(\tau;r,s)\rangle
=
\sum_{\mu=0}^1
\Theta_\mu(\tau;r,s)\,|\mu\rangle.
\end{equation}

The ungauged blocks transform under the modular generators as (use $b=\frac{2r}{k}\bm w_1$ and $c=\frac{2s}{k}\bm w_1$ in Eq. (\ref{SBACK}))
\begin{equation}
\mathbb S:\quad \Theta_\mu\!\left(-\frac1\tau;r,s\right)
=
\frac{(-i\tau)^{1/2}}{\sqrt2}\,
e^{-\,\frac{4\pi i}{k^2}rs}
\sum_{\nu=0}^1
e^{-\pi i\mu\nu}\,
\Theta_\nu(\tau;s,-r),
\label{EHdiscreteS}
\end{equation}
and (see Eq. (\ref{TtransBack}))
\begin{equation}
\mathbb T: \quad \Theta_\mu(\tau+1;r,s)
=
t_\mu\,
e^{-\,\frac{2\pi i}{k^2}r^2}\,
\Theta_\mu(\tau;r,s-r),
\qquad
t_0=1,
\qquad
t_1=i.
\label{EHdiscreteT}
\end{equation}

There are several natural ways to gauge a discrete $\mathbb Z_k$ subgroup of the electric-magnetic $1$-form symmetry in the Eguchi--Hanson construction. The simplest choices are the pure electric and pure magnetic subgroups,
$
L_e=\{(r,0)\mid r\in\mathbb Z_k\},
L_m=\{(0,s)\mid s\in\mathbb Z_k\}.
$
More generally, one may gauge a dyonic subgroup
$
L_m^{\rm dy}=\{(r,mr)\mid r\in\mathbb Z_k\},
$
where the electric and magnetic background labels are correlated. In all cases, the gauged block is obtained by summing over the chosen subgroup, possibly with a quadratic refinement, interpreted as a discrete theta angle, and with an optional Fourier character that projects onto a definite dual sector.

The allowed gaugings are constrained by the mixed electric-magnetic $1$-form anomaly. This anomaly is encoded by the antisymmetric pairing
$
\langle (r,s),(r',s')\rangle
:=
\frac{rs'-sr'}{k}
\in \mathbb Q/\mathbb Z .
$
A subgroup
$
L\subset \mathbb Z_k^{(e)}\times \mathbb Z_k^{(m)}
$
can be gauged only if this pairing is trivial on $L$, namely if
$
\langle \ell,\ell'\rangle=0\,
\text{for all } \ell,\ell'\in L.
$
Equivalently, $L$ must be isotropic (meaning have no anomaly) with respect to the electric-magnetic pairing. The pure electric and pure magnetic subgroups are manifestly isotropic, and so are the cyclic dyonic subgroups $L_m^{\rm dy}=\{(r,mr)\}$. By contrast, the full product
$
\mathbb Z_k^{(e)}\times \mathbb Z_k^{(m)}
$
is not isotropic, and therefore cannot be gauged as an ordinary symmetry without additional anomaly-cancelling structure.

For pure electric gauging, one sums over
\begin{equation}
L_e=\{(r,0)\mid r\in\mathbb Z_k\},
\end{equation}
and defines
\begin{equation}
|\Psi^{({\cal E})}_{a,p_e}(\tau)\rangle
:=
\frac{1}{\sqrt{k}}
\sum_{r\in\mathbb Z_k}
\exp\!\left(
\frac{2\pi i\,p_e}{k}r^2
+\frac{2\pi i}{k}ar
\right)
|\Psi(\tau;r,0)\rangle,
\label{electricgaugedstateEH}
\end{equation}
where $a\in\mathbb Z_k$ labels the Fourier character and $p_e\in\mathbb Z_k$ is the discrete electric theta angle. The linear factor $\exp\!\left(\frac{2\pi i}{k}ar\right)$ is the Fourier character projecting onto a definite dual $\widehat{\mathbb Z_k}$ sector, whereas the quadratic factor $\exp\!\left(\frac{2\pi i\,p_e}{k}r^2\right)$ is the discrete theta angle, or quadratic refinement, intrinsic to the gauging itself. Its origin is as follows. Gauging a $\mathbb Z_k$ $1$-form symmetry means coupling the theory to a $\mathbb Z_k$-valued $2$-form gauge field $B$, and the allowed local topological counterterms are therefore built from $B$. The basic nontrivial one is the Pontryagin-square term \cite{Kapustin:2013qsa}
\begin{equation}
\exp\!\left(\frac{2\pi i\,p_e}{2k}\int_X \mathfrak P(B)\right),
\end{equation}
or, on a spin manifold and for $k$ odd, equivalently $\exp\!\left(\frac{2\pi i\,p_e}{2k}\int_X B\cup B\right)$. To express this in the same convention used in the continuous backgrounds, recall that $\bm w_1$ denotes the fundamental weight of $A_1$, normalized by $\bm w_1^2=(C^{-1})_{11}=\frac12$. The generator associated with the exceptional sphere is then the root $\bm \alpha_1=2\bm w_1$. Writing the discrete background in the weight basis as
\begin{equation}
B_r=\,2r\, \bm w_1,
\end{equation}
one finds
\begin{equation}
\int_X \mathfrak P(B_r)
\equiv
\, (2r)^2 (C^{-1})_{11}
=
\,2r^2
\pmod{2k}.
\end{equation}
 Substituting this into the discrete topological action gives
\begin{equation}
\exp\!\left(\frac{2\pi i\,p_e}{2k}\int_X \mathfrak P(B_r)\right)
=
\exp\!\left(\frac{2\pi i\,p_e}{k}r^2\right),
\end{equation}
which is precisely the quadratic refinement appearing in \eqref{electricgaugedstateEH}. Thus, the phase is not introduced ad hoc to match modular properties; it is the natural local discrete counterterm of the $\mathbb Z_k$ gauging, written here in the same normalization as the continuous backgrounds.

 Expanding in the boundary-sector basis gives
\begin{equation}
|\Psi^{({\cal E})}_{a,p_e}(\tau)\rangle
=
\sum_{\mu=0}^1
\mathcal Z^{({\cal E})}_{\mu;a,p_e}(\tau)\,|\mu\rangle,
\end{equation}
with
\begin{equation}
\mathcal Z^{({\cal E})}_{\mu;a,p_e}(\tau)
=
\frac{1}{\sqrt{k}}
\sum_{r\in\mathbb Z_k}
\exp\!\left(
\frac{2\pi i\,p_e}{k}r^2
+\frac{2\pi i}{k}ar
\right)
\Theta_\mu(\tau;r,0).
\label{electricgaugedblockEH}
\end{equation}
The special choice $p_e=0$ corresponds to a trivial discrete theta angle.

Similarly, pure magnetic gauging is obtained from
\begin{equation}
L_m=\{(0,s)\mid s\in\mathbb Z_k\},
\end{equation}
and gives
\begin{equation}
\mathcal Z^{({\cal M})}_{\mu;b,p_m}(\tau)
=
\frac{1}{\sqrt{k}}
\sum_{s\in\mathbb Z_k}
\exp\!\left(
\frac{2\pi i\,p_m}{k}s^2
+\frac{2\pi i}{k}bs
\right)
\Theta_\mu(\tau;0,s),
\label{magneticgaugedblockEH}
\end{equation}
where $b\in\mathbb Z_k$ is the magnetic Fourier label and $p_m\in\mathbb Z_k$ is the magnetic discrete theta angle. Finally, the diagonal dyonic subgroup
\begin{equation}
L_d=\{(r,-r)\mid r\in\mathbb Z_k\}
\end{equation}
leads to
\begin{equation}
\mathcal Z^{({\cal D})}_{\mu;c,p_d}(\tau)
=
\frac{1}{\sqrt{k}}
\sum_{r\in\mathbb Z_k}
\exp\!\left(
\frac{2\pi i\,p_d}{k}r^2
-\frac{ 2\pi i r^2}{k^2}+\frac{2\pi i}{k}cr
\right)
\Theta_\mu(\tau;r,-r),
\label{dyonicgaugedblockEH}
\end{equation}
where the extra factor
\begin{equation}
\exp\!\left(-\frac{2\pi i}{k^2}r^2\right)
\end{equation}
is precisely the factor in \eqref{EHdiscreteT}.

The modular action on the gauged blocks follows directly from \eqref{EHdiscreteS} and \eqref{EHdiscreteT}. Under $\mathbb S$, pure electric gauging is mapped to pure magnetic gauging:
\begin{equation}
\mathcal Z^{({\cal E})}_{\mu;a,p_e}\!\left(-\frac1\tau\right)
=
\frac{(-i\tau)^{1/2}}{\sqrt2}
\sum_{\nu=0}^1
e^{-\pi i\mu\nu}\,
\mathcal Z^{({\cal M})}_{\nu;-a,p_e}(\tau),
\label{SonelectricgaugedEH}
\end{equation}
so $\mathbb S$ exchanges electric and magnetic gauging, as expected from electric-magnetic duality. By contrast, under $\mathbb T$ one finds
\begin{equation}
\mathcal Z^{({\cal E})}_{\mu;a,p_e}(\tau+1)
=
t_\mu\,
\frac{1}{\sqrt{k}}
\sum_{r\in\mathbb Z_k}
\exp\!\left(
\frac{2\pi i\,p_e}{k}r^2
+\frac{2\pi i}{k}ar
-\frac{2\pi i}{k^2}r^2
\right)
\Theta_\mu(\tau;r,-r),
\label{TonelectricgaugedEH}
\end{equation}
so $\mathbb T$ sends pure electric gauging to a dyonic gauging along the diagonal subgroup. 
In addition, one can show that 
\begin{eqnarray}\nonumber
(\mathbb S \mathbb T^{-1})^3\mathcal Z^{({{\cal E}})}_{\mu;a,p_e}=\mathcal Z^{({\cal E})}_{\mu;-a,p_e}\,, \quad (\mathbb S \mathbb T^{-1})^3\mathcal Z^{({{\cal M}})}_{\mu;b,p_m}=\mathcal Z^{({\cal M})}_{\mu;-b,p_m}\,,\quad (\mathbb S \mathbb T^{-1})^3\mathcal Z^{({{\cal D}})}_{\mu;c,p_d}=\mathcal Z^{({\cal D})}_{\mu;-c,p_d}\,,\\
\end{eqnarray}
and thus, as expected, $(\mathbb S \mathbb T^{-1})^3\equiv \mathbb C$ is the charge conjugation operator, and we used the fact that for the $A_1$ algebra, $\mu=-\mu$ mod $2$. 
Thus, a pure electric $\mathbb Z_k$-gauged Eguchi--Hanson block is not closed under the full duality group generated by $\mathbb S$ and $\mathbb T$, since $\mathbb S$ maps it to a magnetic gauging and $\mathbb T$ maps it to a dyonic gauging. However, it is closed under the particular element $(\mathbb S\mathbb T^{-1})^3$, which acts simply as charge conjugation on the discrete character label.

\subsection{Gluing Eguchi--Hanson blocks after $\mathbb Z_k$ $1$-form gauging}

Finally, let us examine the gluing of the pure electric $\mathbb Z_k$-gauged states on the two Eguchi--Hanson halves. As explained at the end of Section \ref{Gluing ALE spaces with 1-form backgrounds}, when $k$ is even, a $\mathbb Z_k$ $2$-form background has an additional $\mathbb Z_2$ boundary data on the lens-space interface. Treating this case properly requires a refined prescription. We therefore restrict here to odd $k$, for which
$
H^2(S^3/\mathbb Z_2,\mathbb Z_k)=0
$
and no extra discrete boundary matching condition arises. In this case, the $\mathbb Z_k$ labels on the two Eguchi--Hanson pieces give a complete parametrization of the discrete electric backgrounds on the glued manifold.

Repeating the analysis of Section \ref{Gluing ALE spaces with 1-form backgrounds}, we find
\begin{eqnarray}\nonumber
Z^{({\cal E})}_{\rm glue}
:=
\big\langle
\Psi^{({\cal E})}_{L;a_L,p_L}(\bar\tau)
\big|\,
\Psi^{({\cal E})}_{R;a_R,p_R}(\tau)
\big\rangle
&=&
\frac1k
\sum_{r_R,r_L\in\mathbb Z_k}
\exp\!\left[
\frac{2\pi i}{k}
\left(
p_Rr_R^2-p_Lr_L^2+a_Rr_R-a_Lr_L
\right)
\right]
\\
&&\times
Z_{\rm glue}\!\left(
\tau,\bar\tau;
\frac{2r_R}{k},\frac{2r_L}{k};0,0
\right),
\end{eqnarray}
where $Z_{\rm glue}$ is the gluing partition function with electric and magnetic backgrounds, given in \eqref{eq:S2S2-full-background}.
The electric background variables on the glued $S^2\times S^2$ are related to the Eguchi--Hanson labels by
\begin{equation}
x_1=\frac{r_L-r_R}{k},
\qquad
x_2=-\,\frac{r_R+r_L}{k}.
\end{equation}
For odd $k$, this change of variables is invertible over $\mathbb Z_k$. Thus, the sum over $(r_R,r_L)$ may be replaced by a sum over all discrete electric background sectors $(kx_1,kx_2)\in\mathbb Z_k^2$ of the closed manifold. Substituting this change of variables into the phase gives
\begin{equation}
Z^{({\cal E})}_{\rm glue}
=
Z_{S^2\times S^2}^{({\cal E})}
=
\frac1k
\sum_{kx_1,kx_2\in\mathbb Z_k}
\exp\!\left[
\frac{2\pi i}{k}\,
\Phi(x_1,x_2)
\right]
Z_{\rm glue}\!\left(
\tau,\bar\tau;
x_1,x_2;0,0
\right),
\end{equation}
where
\begin{equation}
\Phi(x_1,x_2)
=
2^{-2}
\left[
p_R k^2(x_1+x_2)^2
-
p_L k^2(x_1-x_2)^2
\right]
-
2^{-1}
\left[
(a_R+a_L)kx_1
+
(a_R-a_L)kx_2
\right].
\end{equation}
Here, all quantities in the exponent are understood modulo $k$ in $\mathbb Z_k$.

The parameters $p_L,p_R$ and $a_L,a_R$ belong to the two gauged Eguchi--Hanson states before gluing. After gluing, they appear only through the combinations that weight the closed-manifold background sum. Thus, the most general gluing of two independently weighted Eguchi-Hanson states should not be identified directly with the standard $\mathbb Z_k$ gauging of Maxwell theory on $S^2\times S^2$; it is a more general gluing with independent boundary-state weights. To obtain the natural closed-manifold gauging, one should choose compatible left and right discrete theta angles and characters. In particular, for the symmetric choice
\begin{equation}
p_L=p_R=:p,
\qquad
a_L=a_R=:a,
\end{equation}
the phase reduces to
\begin{equation}
\Phi(x_1,x_2)
=
p\,k^2x_1x_2-a\,kx_1.
\end{equation}
The quadratic term is then proportional to the intersection pairing on $S^2\times S^2$. Thus, the symmetric gluing produces the expected closed-manifold discrete theta weight, while the more general choices of $(p_L,p_R,a_L,a_R)$ describe a more general weighting of the two Eguchi-Hanson boundary states.

{\bf \flushleft{Acknowledgments:}} I am grateful to Erich Poppitz and Yuya Tanizaki for comments on the manuscript. ChatGPT by OpenAI was used as an aid in editing and checking intermediate steps. The author takes full responsibility for the correctness of the results. This work was supported by the STFC under grant ST/X000591/1.

\appendix

\section{$S$-duality action on the path integral defined on a manifold with boundary}
\label{Sduality on the path integral}

Here we take a pedestrian approach to deriving the action of $S$-duality on a path integral defined on a manifold $M$ with boundary $\partial M$. The role of $S$-duality in the presence of boundaries and walls has been studied extensively; see, for example, \cite{Witten:2003ya,Zucchini:2003in,Kapustin:2009av,Seiberg:2016gmd}. A path integral on $M$ naturally defines a state, or wavefunction, on the boundary $\partial M$. In this work, we adapt the ideas in \cite{Witten:2003ya,Zucchini:2003in,Kapustin:2009av,Seiberg:2016gmd} to determine how this boundary state transforms under $ S$-duality. For simplicity, and in accordance with the setup relevant to us, we neglect possible topological degrees of freedom localized on the boundary. Also, we consider the path integral on the upper half-space of $\mathbb{R}^4$. The resulting analysis can be straightforwardly generalized to more general manifolds. The advantage of this simple geometry is that it makes the relevant physics particularly transparent.

Let
$
M=\{z\geq 0\}\subset \mathbb R^4
$
with Euclidean coordinates
$
(t,x,y,z),
$
where $t$ is Euclidean time. We orient $M$ by
$
dt\wedge dx\wedge dy\wedge dz.
$
The boundary is
$
Y\equiv\partial M=\{z=0\},
$
and we take the orientation on $Y$ such that
$
dt\wedge dx\wedge dy
$
is positive.

Let $A$ be a $U(1)$ gauge field on $M$, with curvature
$
F=dA.
$
We take the Euclidean Maxwell action to be
\begin{equation}
S[A]
=
\frac{1}{g^2}\int_M F\wedge \star F
-
\frac{i\theta}{8\pi^2}\int_M F\wedge F,
\label{eq:Maxwell-action-boundary}
\end{equation}
and define the complex coupling constant $\tau$ as:
\begin{equation}
\tau
=
\frac{\theta}{2\pi}
+
i\,\frac{4\pi}{g^2}.
\label{eq:tau-def-boundary}
\end{equation}

Introduce the Euclidean electric and magnetic fields by
\begin{equation}
E_i=F_{ti},
\qquad i=x,y,z,
\end{equation}
and
\begin{equation}
B_x=F_{yz},
\qquad
B_y=F_{zx},
\qquad
B_z=F_{xy}.
\end{equation}
Thus
\begin{align}
F
&=
E_x\,dt\wedge dx
+
E_y\,dt\wedge dy
+
E_z\,dt\wedge dz
\nonumber\\
&\hspace{2cm}
+
B_x\,dy\wedge dz
+
B_y\,dz\wedge dx
+
B_z\,dx\wedge dy,
\end{align}
and hence\footnote{With our orientation convention,
\begin{eqnarray}\nonumber
\star(dt\wedge dx)=dy\wedge dz,\quad
\star(dy\wedge dz)=dt\wedge dx,\quad
\star(dt\wedge dy)=dz\wedge dx,\\
\star(dz\wedge dx)=dt\wedge dy,\quad
\star(dt\wedge dz)=dx\wedge dy,\quad
\star(dx\wedge dy)=dt\wedge dz.
\end{eqnarray}.}
\begin{align}
\star F
&=
B_x\,dt\wedge dx
+
B_y\,dt\wedge dy
+
B_z\,dt\wedge dz
\nonumber\\
&\hspace{2cm}
+
E_x\,dy\wedge dz
+
E_y\,dz\wedge dx
+
E_z\,dx\wedge dy.
\end{align}
In other words, in these conventions,
$
\star(E,B)=(B,E).
$
It follows that
\begin{equation}
F\wedge \star F
=
(E^2+B^2)\,dt\,dx\,dy\,dz,
\end{equation}
and
\begin{equation}
F\wedge F
=
2E\cdot B\,dt\,dx\,dy\,dz.
\end{equation}
Therefore the action \eqref{eq:Maxwell-action-boundary} becomes
\begin{equation}
S[A]
=
\int_M dt\,dx\,dy\,dz
\left[
\frac{1}{g^2}(E^2+B^2)
-
\frac{i\theta}{4\pi^2}E\cdot B
\right].
\label{eq:action-EB}
\end{equation}

The ordinary Hamiltonian momenta with respect to Euclidean time $t$ are conjugate to
$
A_x, A_y, A_z.
$
They are
\begin{equation}
\pi_i
=
\frac{\partial \mathcal L}{\partial(\partial_t A_i)}
=
\frac{2}{g^2}E_i
-
\frac{i\theta}{4\pi^2}B_i.
\label{eq:time-canonical-momentum}
\end{equation}
The component $A_t$ is not a Hamiltonian coordinate; it acts as a Lagrange multiplier imposing Gauss's law.

\subsection*{Boundary variation and boundary canonical data}

It is useful to define the $2$-form
\begin{equation}
G
=
\frac{2}{g^2}\star F
-
\frac{i\theta}{4\pi^2}F.
\label{eq:G-def-boundary}
\end{equation}
Then
\begin{align}
\delta S
&=
\frac{2}{g^2}\int_M \delta F\wedge \star F
-
\frac{i\theta}{4\pi^2}\int_M \delta F\wedge F
\nonumber\\
&=
\int_M d(\delta A)\wedge G.
\end{align}
Using
$
d(\delta A\wedge G)
=
d(\delta A)\wedge G
-
\delta A\wedge dG,
$
we obtain
\begin{equation}
\delta S
=
\int_M \delta A\wedge dG
+
\int_Y \delta A_\parallel\wedge G_\parallel.
\label{eq:variation-with-boundary}
\end{equation}
Thus the bulk equation of motion is
\begin{equation}
dG=0.
\end{equation}

On the boundary we write the pullback of $A$ as
\begin{equation}
a=A_\parallel
=
a_t\,dt+a_x\,dx+a_y\,dy.
\end{equation}
The Dirichlet boundary condition is
\begin{equation}
A_\parallel|_Y=a.
\end{equation}
This fixes only the tangential components of the gauge field:
$
A_t|_Y=a_t,
A_x|_Y=a_x,
A_y|_Y=a_y.
$
It does not fix $A_z$.

The tangential field strength on the boundary is fixed by
\begin{equation}
F_\parallel=da.
\end{equation}
Writing
\begin{equation}
da
=
e_x\,dt\wedge dx
+
e_y\,dt\wedge dy
+
b_z\,dx\wedge dy,
\end{equation}
we have
\begin{eqnarray}
e_x=\partial_t a_x-\partial_x a_t,\quad
e_y=\partial_t a_y-\partial_y a_t,
\quad
b_z=\partial_x a_y-\partial_y a_x.
\end{eqnarray}
Thus the Dirichlet boundary condition fixes
\begin{equation}
E_x|_Y=e_x,
\qquad
E_y|_Y=e_y,
\qquad
B_z|_Y=b_z.
\end{equation}
It does not fix the normal components
\begin{equation}
E_z|_Y,
\qquad
B_x|_Y,
\qquad
B_y|_Y.\label{normalc}
\end{equation}

The pullback of $G$ to the boundary has components
\begin{eqnarray}
G_{tx}
=
\frac{2}{g^2}B_x
-
\frac{i\theta}{4\pi^2}E_x,\quad
G_{ty}
=
\frac{2}{g^2}B_y
-
\frac{i\theta}{4\pi^2}E_y,\quad
G_{xy}
=
\frac{2}{g^2}E_z
-
\frac{i\theta}{4\pi^2}B_z.
\end{eqnarray}
Therefore
\begin{align}
\delta a\wedge G_\parallel
&=
\left(
\delta a_t\,G_{xy}
-
\delta a_x\,G_{ty}
+
\delta a_y\,G_{tx}
\right)
dt\wedge dx\wedge dy.
\end{align}
Thus the boundary contribution to the variation can be written as
\begin{equation}
\delta S_{\partial M}
=
\int_Y dt\,dx\,dy
\left(
\delta a_t\,p_t
+
\delta a_x\,p_x
+
\delta a_y\,p_y
\right),
\end{equation}
where
\begin{align}\nonumber
p_t
&=
G_{xy}
=
\frac{2}{g^2}E_z
-
\frac{i\theta}{4\pi^2}B_z,
\\\nonumber
p_x
&=
-G_{ty}
=
-\frac{2}{g^2}B_y
+
\frac{i\theta}{4\pi^2}E_y,
\\
p_y
&=
G_{tx}
=
\frac{2}{g^2}B_x
-
\frac{i\theta}{4\pi^2}E_x.
\label{eq:py-original-fields}
\end{align}
These $p_\alpha$, with $\alpha=t,x,y$, are the momenta conjugate to the boundary values $a_\alpha$ across the wall $z=0$. They should not be confused with the ordinary time-Hamiltonian momenta $\pi_i$ in \eqref{eq:time-canonical-momentum}.

The path integral with the Dirichlet boundary condition defines a boundary wavefunctional
\begin{equation}
\Psi_\tau[a]
=
\int_{A_\parallel=a}\mathcal D A\,
e^{-S_\tau[A]}.
\label{eq:boundary-wavefunctional}
\end{equation}

\subsection*{First-order formulation and the dual field}

To derive the dual theory, introduce  a dual gauge field $A_D$, with
$
F_D=dA_D.
$
We use the parent action
\begin{equation}
S_{\rm par}[F,A_D]
=
S_0[F]
+
\frac{i}{2\pi}\int_M F\wedge F_D,
\label{eq:parent-action-boundary}
\end{equation}
where
\begin{equation}
S_0[F]
=
\frac{1}{g^2}\int_M F\wedge \star F
-
\frac{i\theta}{8\pi^2}\int_M F\wedge F.
\end{equation}
Define the dual electric and magnetic fields by
\begin{equation}
E_{D,i}=(F_D)_{ti},
\end{equation}
and
\begin{equation}
B_{D,x}=(F_D)_{yz},
\qquad
B_{D,y}=(F_D)_{zx},
\qquad
B_{D,z}=(F_D)_{xy}.
\end{equation}
Then
\begin{equation}
F\wedge F_D
=
(E\cdot B_D+B\cdot E_D)\,dt\,dx\,dy\,dz.
\end{equation}
Hence the parent Lagrangian density is
\begin{equation}
\mathcal L_{\rm par}
=
\frac{1}{g^2}(E^2+B^2)
-
\frac{i\theta}{4\pi^2}E\cdot B
+
\frac{i}{2\pi}
\left(
E\cdot B_D+B\cdot E_D
\right).
\end{equation}

Varying with respect to $E_i$ gives
\begin{equation}
0
=
\frac{2}{g^2}E_i
-
\frac{i\theta}{4\pi^2}B_i
+
\frac{i}{2\pi}B_{D,i}.
\end{equation}
Therefore
\begin{equation}
B_D
=
\frac{\theta}{2\pi}B
+
i\,\frac{4\pi}{g^2}E.
\label{eq:BD-map}
\end{equation}
Equivalently, using the Hamiltonian momentum $\pi_i$ in \eqref{eq:time-canonical-momentum},
\begin{equation}
B_D=2\pi i\,\pi,
\qquad
\pi=-\frac{i}{2\pi}B_D.
\label{eq:BD-pi-map}
\end{equation}

Similarly, varying with respect to $B_i$ gives
\begin{equation}
0
=
\frac{2}{g^2}B_i
-
\frac{i\theta}{4\pi^2}E_i
+
\frac{i}{2\pi}E_{D,i}.
\end{equation}
Therefore
\begin{equation}
E_D
=
\frac{\theta}{2\pi}E
+
i\,\frac{4\pi}{g^2}B.
\label{eq:ED-map}
\end{equation}
Equations \eqref{eq:BD-map} and \eqref{eq:ED-map} are the component form of the bulk $S$-duality map.

They imply
\begin{equation}
G=-\frac{i}{2\pi}F_D.
\label{eq:G-dual-relation}
\end{equation}
Indeed, the $dt\wedge dx^i$ components of $G$ are
$
G_{ti}
=
\frac{2}{g^2}B_i
-
\frac{i\theta}{4\pi^2}E_i
=
-\frac{i}{2\pi}E_{D,i},
$
and the purely spatial components similarly give
$
\frac{2}{g^2}E
-
\frac{i\theta}{4\pi^2}B
=
-\frac{i}{2\pi}B_D.
$

In terms of the self-dual and anti-self-dual parts
$
F^\pm=\frac12(F\pm \star F),
$
the original action can also be written as
\begin{equation}
S_0[F]
=
-\frac{i}{4\pi}
\int_M
\left(
\tau F^+\wedge F^+
+
\bar\tau F^-\wedge F^-
\right).
\end{equation}
The equation of motion obtained by varying $F$ in the parent action gives
\begin{equation}
F_D^+=\tau F^+,
\qquad
F_D^-=\bar\tau F^-.
\end{equation}
Substituting back into the parent action gives the dual Maxwell action
\begin{equation}
S_{\rm dual}[A_D]
=
\frac{i}{4\pi}\int_M
\left(
\frac{1}{\tau}F_D^+\wedge F_D^+
+
\frac{1}{\bar\tau}F_D^-\wedge F_D^-
\right).
\end{equation}
This has the same form as the original action with
\begin{equation}
\tau_D=-\frac1{\tau}.
\end{equation}

\subsection*{Boundary Fourier kernel}

Restricting \eqref{eq:G-dual-relation} to the boundary gives
\begin{equation}
G_\parallel
=
-\frac{i}{2\pi}(F_D)_\parallel.
\end{equation}
If
\begin{equation}
a_D=(A_D)_\parallel
=
a_{D,t}\,dt+a_{D,x}\,dx+a_{D,y}\,dy,
\end{equation}
then
\begin{equation}
(F_D)_\parallel=da_D.
\end{equation}
Write
\begin{equation}
da_D
=
e_{D,x}\,dt\wedge dx
+
e_{D,y}\,dt\wedge dy
+
b_{D,z}\,dx\wedge dy,
\end{equation}
where
\begin{eqnarray}
e_{D,x}
=
\partial_t a_{D,x}-\partial_x a_{D,t},\quad
e_{D,y}
=
\partial_t a_{D,y}-\partial_y a_{D,t},\quad
b_{D,z}
=
\partial_x a_{D,y}-\partial_y a_{D,x}.
\end{eqnarray}
Then
\begin{equation}
G_\parallel=-\frac{i}{2\pi}da_D.
\end{equation}
Using the definitions of $p_t,p_x,p_y$, this gives
\begin{align}\nonumber
p_t
&=
-\frac{i}{2\pi}b_{D,z},
\\\nonumber
p_x
&=
+\frac{i}{2\pi}e_{D,y},
\\
p_y
&=
-\frac{i}{2\pi}e_{D,x}.
\label{eq:py-dual}
\end{align}
Thus the dual boundary field $a_D$ parametrizes the momentum conjugate to $a$.

Now consider the mixed term in the parent action:
$
\frac{i}{2\pi}\int_M F\wedge F_D
=
\frac{i}{2\pi}\int_M F\wedge dA_D.
$
Using
$
d(F\wedge A_D)
=
dF\wedge A_D+F\wedge dA_D,
$
we have
\begin{equation}
\int_M F\wedge dA_D
=
\int_Y F_\parallel\wedge a_D
-
\int_M dF\wedge A_D.
\end{equation}
With the Dirichlet condition $A_\parallel=a$, we have $F_\parallel=da$, so the boundary contribution to the parent action is
\begin{equation}
S_{\partial,\rm par}
=
\frac{i}{2\pi}\int_Y da\wedge a_D.
\end{equation}
Since
$
d(a\wedge a_D)=da\wedge a_D-a\wedge da_D,
$
and assuming $Y$ has no boundary, we get
\begin{equation}
\int_Y da\wedge a_D
=
\int_Y a\wedge da_D.
\end{equation}
Therefore
\begin{equation}
S_{\partial,\rm par}
=
\frac{i}{2\pi}\int_Y a\wedge da_D.
\end{equation}
Because the Euclidean path integral contains $e^{-S}$, this boundary term contributes the kernel
\begin{equation}
\exp\left[
-\frac{i}{2\pi}\int_Y a\wedge da_D
\right].
\label{eq:boundary-kernel-form}
\end{equation}

In components,
\begin{align}
a\wedge da_D
&=
(a_t\,dt+a_x\,dx+a_y\,dy)
\nonumber\\
&\quad\wedge
\left(
e_{D,x}\,dt\wedge dx
+
e_{D,y}\,dt\wedge dy
+
b_{D,z}\,dx\wedge dy
\right)
\nonumber\\
&=
\left(
a_t b_{D,z}
-
a_x e_{D,y}
+
a_y e_{D,x}
\right)
dt\wedge dx\wedge dy.
\end{align}
Hence
\begin{equation}
\exp\left[
-\frac{i}{2\pi}\int_Y a\wedge da_D
\right]
=
\exp\left[
-\frac{i}{2\pi}
\int_Y dt\,dx\,dy
\left(
a_t b_{D,z}
-
a_x e_{D,y}
+
a_y e_{D,x}
\right)
\right].
\label{eq:boundary-kernel-components}
\end{equation}
Using \eqref{eq:py-dual}, the same exponent can be written as
$
\int_Y dt\,dx\,dy
\left(
a_t p_t+a_xp_x+a_yp_y
\right).
$
Thus
\begin{equation}
\exp\left[
-\frac{i}{2\pi}\int_Y a\wedge da_D
\right]
=
\exp\left[
\int_Y dt\,dx\,dy\,
a_\alpha p_\alpha
\right],
\qquad
\alpha=t,x,y.\label{importantpa}
\end{equation}

The parent-action derivation above implements the standard representative
$\mathbb S_{\rm std}$ of the transformation $\tau\mapsto-1/\tau$, for which
$F\mapsto F_D$ and the boundary kernel has the sign displayed in
\eqref{eq:boundary-kernel-form}.  The active convention used in the main text
differs by charge conjugation,
\begin{equation}
\mathbb S=\mathbb C\circ\mathbb S_{\rm std},
\qquad
F\longmapsto-F_D,
\qquad
F_D\longmapsto F.
\end{equation}
It is implemented on the dual integration variable by
\begin{equation}
A_D^{\rm act}:=-A_D,
\qquad
a_D^{\rm act}:=-a_D.
\end{equation}
Substituting $a_D=-a_D^{\rm act}$ in the standard Fourier kernel reverses its
sign. Dropping the superscript ``act'' from this point onward, the boundary
wavefunctional therefore transforms under the active operation $\mathbb S$ as
\begin{equation}
\Psi_{-1/\tau}[a_D]
=
\int \mathcal D a\,
\exp\left[
\frac{i}{2\pi}\int_Y a\wedge da_D
\right]
\Psi_\tau[a],
\label{eq:S-duality-boundary-wavefunctional}
\end{equation}
up to normalization. In components,
\begin{equation}
\Psi_{-1/\tau}[a_D]
=
\int \mathcal D a\,
\exp\left[
\frac{i}{2\pi}
\int_Y dt\,dx\,dy
\left(
a_t b_{D,z}
-
a_x e_{D,y}
+
a_y e_{D,x}
\right)
\right]
\Psi_\tau[a].
\end{equation}

This equation should be interpreted as a change of boundary polarization. The original wavefunctional $\Psi_\tau[a]$ is computed with $a$ fixed. The subsequent integral over $a$ is the Fourier transform from the $a$-polarization to the conjugate polarization parametrized by $a_D$. Further, from (\ref{normalc}, \ref{eq:py-original-fields}, \ref{importantpa}), we see that the Fourier integral variable $a$ is conjugate to $e_z,b_x,b_y$, the unfixed components of the field strength at the boundary (at $\theta=0$), as one should expect.  

\subsection*{Flat boundary bundles and torsion holonomy}

The preceding formulas are local differential-form expressions. They are sufficient when the boundary gauge fields $a$ and $a_D$ are globally defined 
$1$-forms. However, for genuine $U(1)$ gauge fields, the boundary data should be regarded as a connection on a line bundle over $Y$, not merely as a globally defined $1$-form.

This distinction becomes important when the boundary connection is flat but topologically nontrivial. Suppose
\begin{equation}
F_\parallel=da=0
\end{equation}
as a differential form, but the boundary line bundle has nontrivial holonomy. Then $a$ is locally pure gauge, but it need not be globally gauge-trivial. Flat boundary $U(1)$ connections are classified by
\begin{equation}
\operatorname{Hom}(H_1(Y;\mathbb Z),U(1)).
\end{equation}
If $H_1(Y;\mathbb Z)$ has torsion, then there can be finite-order flat holonomies.

In this case, the naive kernel
$
\exp\left[
\frac{i}{2\pi}\int_Y a\wedge da_D
\right]
$
is incomplete. If both $a$ and $a_D$ are flat torsion connections, then the differential forms $da$ and $da_D$ vanish, and the integral expression appears to give $1$. This misses the global pairing between torsion sectors.

The correct replacement is a differential-cohomology pairing. Schematically, if $\check a$ and $\check a_D$ denote the differential cohomology classes of the two boundary connections, then the local expression should be replaced by
\begin{equation}
\exp\left[
2\pi i\int_Y \check a\cup \check a_D
\right].
\label{eq:differential-cohomology-kernel}
\end{equation}
When the connections are topologically trivial, \eqref{eq:differential-cohomology-kernel} reduces to the differential-form expression. 
For spaces with torsion, the torsion part of the $S$-duality kernel is then
\begin{equation}
K_{\rm tors}(\alpha,\beta)
=
\exp\left[2\pi i\,\lambda(\alpha,\beta)\right],
\label{eq:torsion-kernel}
\end{equation}
where $\alpha$ and $\beta$ play the roles of $a$ and $a_D$, and $\lambda(\alpha,\beta)$ is their pairing.
Thus, the full boundary transform includes both a continuous functional Fourier transform over the de Rham part of the connection and a finite Fourier transform over the torsion sectors.

In particular, if the bulk line bundle restricts to a flat torsion bundle on the boundary, then
$
F_\parallel=0
$
as a differential form, but the boundary condition still contains nontrivial global data. This global data is not seen by the local expression $a\wedge da_D$, but it is seen by the torsion linking pairing \eqref{eq:torsion-kernel}, which is basically a discrete Fourier transform.

\section{Cohomology of Eguchi--Hanson space and discrete $2$-form backgrounds}
\label{app:EH-cohomology-backgrounds}

In this appendix, we collect some elementary topological facts about Eguchi--Hanson space and explain how they enter the discussion of continuous $U(1)$ and discrete $\mathbb Z_k$ $2$-form backgrounds. The main point is that the bulk Eguchi--Hanson space contains one compact $2$-cycle, while its asymptotic boundary is the lens space $S^3/\mathbb Z_2$. Consequently, continuous $U(1)$ $2$-form backgrounds have one bulk period, whereas discrete $\mathbb Z_k$ backgrounds may also have a residual boundary torsion class when $k$ is even.

Topologically, Eguchi--Hanson space $X_{\rm EH}$ is the disk bundle over $S^2$. Its ordinary integral homology is
\begin{eqnarray}\nonumber
&&H_0(X_{\rm EH},\mathbb Z)\cong \mathbb Z,
\qquad
H_1(X_{\rm EH},\mathbb Z)=0,
\qquad
H_2(X_{\rm EH},\mathbb Z)\cong \mathbb Z,
\qquad
H_3(X_{\rm EH},\mathbb Z)=0,\\
&&H_4(X_{\rm EH},\mathbb Z)=0.
\end{eqnarray}
The vanishing of $H_4(X_{\rm EH},\mathbb Z)$ is a consequence of noncompactness. The Betti numbers are therefore
\begin{equation}
b_0=1,
\qquad
b_1=0,
\qquad
b_2=1,
\qquad
b_3=0,
\qquad
b_4=0,
\end{equation}
and hence the Euler characteristic is
\begin{equation}
\chi(X_{\rm EH})=b_0+b_2=2.
\end{equation}
The exceptional sphere has self-intersection $2$, so, in our convention, the intersection form is positive definite. Thus, in that orientation, the signature is
\begin{equation}
\sigma(X_{\rm EH})=1.
\end{equation}
With the opposite orientation, the sign of the signature is reversed.

The asymptotic boundary of the Eguchi--Hanson space is
\begin{equation}
Y:=\partial X_{\rm EH}=S^3/\mathbb Z_2\simeq \mathbb{RP}^3.
\end{equation}
Its integral homology is
\begin{equation}
H_0(Y,\mathbb Z)\cong \mathbb Z,
\qquad
H_1(Y,\mathbb Z)\cong \mathbb Z_2,
\qquad
H_2(Y,\mathbb Z)=0,
\qquad
H_3(Y,\mathbb Z)\cong \mathbb Z.
\end{equation}
Thus $Y$ has no ordinary $2$-cycles, but it does have a torsion $1$-cycle. This torsion $1$-cycle is responsible for the $\mathbb Z_2$ flat holonomy sectors at infinity.

Let us consider continuous $U(1)$ $2$-form backgrounds. The relevant cohomology group is
\begin{equation}
H^2(X_{\rm EH},U(1)).
\end{equation}
By the universal coefficient theorem \cite{hatcher2002algebraic},
\begin{equation}
0\longrightarrow
\mathrm{Ext}\!\left(H_1(X_{\rm EH},\mathbb Z),U(1)\right)
\longrightarrow
H^2(X_{\rm EH},U(1))
\longrightarrow
\mathrm{Hom}\!\left(H_2(X_{\rm EH},\mathbb Z),U(1)\right)
\longrightarrow 0\,,
\end{equation}
using
\begin{equation}
H_1(X_{\rm EH},\mathbb Z)=0,
\qquad
H_2(X_{\rm EH},\mathbb Z)\cong \mathbb Z,
\end{equation}
we obtain
\begin{equation}
H^2(X_{\rm EH},U(1))
\cong
\mathrm{Hom}(\mathbb Z,U(1))
\cong
U(1).
\end{equation}
Thus, a continuous $U(1)$ $2$-form background has one independent bulk period, measured on the exceptional sphere. Equivalently, if $B$ is the background $2$-form connection, then its gauge-invariant surface holonomy is
\begin{equation}
\exp\!\left(i\int_{\Sigma} B\right)\in U(1),
\end{equation}
or, in additive notation,
\begin{equation}
\frac{1}{2\pi}\int_{\Sigma}B\in \mathbb R/\mathbb Z.
\end{equation}
In the normalization used in the main text, where the $A_1$ root is $\bm \alpha_1=2\bm w_1$, the corresponding background coordinate has period $2$:
\begin{equation}
b\sim b+2.
\end{equation}
Thus the continuous background parameter may be viewed as an element of
\begin{equation}
\mathbb R/2\mathbb Z.
\end{equation}

The boundary behavior is different. For the boundary lens space, the universal coefficient theorem gives
\begin{equation}
0\longrightarrow
\mathrm{Ext}\!\left(H_1(Y,\mathbb Z),U(1)\right)
\longrightarrow
H^2(Y,U(1))
\longrightarrow
\mathrm{Hom}\!\left(H_2(Y,\mathbb Z),U(1)\right)
\longrightarrow 0.
\end{equation}
Since
\begin{equation}
H_1(Y,\mathbb Z)\cong \mathbb Z_2,
\qquad
H_2(Y,\mathbb Z)=0,
\end{equation}
and\footnote{Using the  quotienting definition of the extension $\mathrm{Ext}(\mathbb Z_n,A)\cong A/nA$, we have $\mathrm{Ext}(\mathbb Z_2,U(1))\cong U(1)/2U(1)\cong 0$. }
\begin{equation}
\mathrm{Ext}(\mathbb Z_2,U(1))=0,
\end{equation}
therefore
\begin{equation}
H^2(Y,U(1))=0.
\end{equation}
Thus, a continuous $U(1)$ $2$-form background has no independent continuous boundary holonomy on $Y$, in agreement with the fact that $Y$ has no ordinary $2$-cycles. This should be distinguished from an ordinary $U(1)$ gauge field, whose flat boundary holonomies are classified by
\begin{equation}
H^1(Y,U(1))
\cong
\mathrm{Hom}(H_1(Y,\mathbb Z),U(1))
\cong
\mathrm{Hom}(\mathbb Z_2,U(1))
\cong
\mathbb Z_2.
\end{equation}

We now restrict the continuous $U(1)$ $1$-form symmetry to a discrete subgroup $\mathbb Z_k$. The corresponding background is a $\mathbb Z_k$-valued $2$-form gauge field, with bulk sectors classified by
\begin{equation}
H^2(X_{\rm EH},\mathbb Z_k).
\end{equation}
Again using the universal coefficient theorem,
\begin{equation}
0\longrightarrow
\mathrm{Ext}\!\left(H_1(X_{\rm EH},\mathbb Z),\mathbb Z_k\right)
\longrightarrow
H^2(X_{\rm EH},\mathbb Z_k)
\longrightarrow
\mathrm{Hom}\!\left(H_2(X_{\rm EH},\mathbb Z),\mathbb Z_k\right)
\longrightarrow 0.
\end{equation}
Since $H_1(X_{\rm EH},\mathbb Z)=0$ and $H_2(X_{\rm EH},\mathbb Z)\cong \mathbb Z$, we find
\begin{equation}
H^2(X_{\rm EH},\mathbb Z_k)
\cong
\mathrm{Hom}(\mathbb Z,\mathbb Z_k)
\cong
\mathbb Z_k.
\end{equation}
Thus, there are $k$ bulk $\mathbb Z_k$ $2$-form background sectors. However, one must distinguish between the odd and even $k$. 

The distinction appears when one restricts the discrete background to the boundary. The boundary class lies in
\begin{equation}
H^2(Y,\mathbb Z_k)
=
H^2(S^3/\mathbb Z_2,\mathbb Z_k).
\end{equation}
Using the universal coefficient theorem,
\begin{equation}
0\longrightarrow
\mathrm{Ext}\!\left(H_1(Y,\mathbb Z),\mathbb Z_k\right)
\longrightarrow
H^2(Y,\mathbb Z_k)
\longrightarrow
\mathrm{Hom}\!\left(H_2(Y,\mathbb Z),\mathbb Z_k\right)
\longrightarrow 0.
\end{equation}
Since $H_2(Y,\mathbb Z)=0$ and $H_1(Y,\mathbb Z)\cong \mathbb Z_2$, this reduces to
\begin{equation}
H^2(Y,\mathbb Z_k)
\cong
\mathrm{Ext}(\mathbb Z_2,\mathbb Z_k).
\end{equation}
To compute this Ext group, use the quotienting
\begin{equation}
\mathrm{Ext}(\mathbb Z_n,A)\cong A/nA.
\end{equation}
Therefore
\begin{equation}
\mathrm{Ext}(\mathbb Z_2,\mathbb Z_k)
\cong
\mathbb Z_k/2\mathbb Z_k.
\end{equation}
If $k$ is odd, multiplication by $2$ is invertible modulo $k$, and hence
\begin{equation}
2\mathbb Z_k=\mathbb Z_k,
\qquad
\mathbb Z_k/2\mathbb Z_k=0.
\end{equation}
If $k$ is even, then
\begin{equation}
2\mathbb Z_k=\{0,2,4,\ldots,k-2\},
\end{equation}
and the quotient is
\begin{equation}
\mathbb Z_k/2\mathbb Z_k\cong \mathbb Z_2.
\end{equation}
Thus
\begin{equation}
H^2(S^3/\mathbb Z_2,\mathbb Z_k)
\cong
\mathbb Z_{\gcd(2,k)}
=
\begin{cases}
0, & k \ \mathrm{odd},\\[2mm]
\mathbb Z_2, & k \ \mathrm{even}.
\end{cases}
\end{equation}

This observation is useful when gluing two Eguchi--Hanson spaces. Let
\begin{equation}
M=S^2\times S^2=X_R\cup_Y X_L,
\qquad
Y=S^3/\mathbb Z_2,
\end{equation}
where $X_R$ and $X_L$ are the two Eguchi--Hanson pieces. For fixed $\mathbb Z_k$ electric backgrounds on the two pieces, labeled by
\begin{equation}
r_R,r_L\in \mathbb Z_k,
\end{equation}
the induced closed-manifold labels may be written as
\begin{equation}
x_1=r_L-r_R,
\qquad
x_2=-r_R-r_L
\qquad
\mathrm{mod}\ k.
\end{equation}
Equivalently,
\begin{equation}
\begin{pmatrix}
x_1\\
x_2
\end{pmatrix}
=
\begin{pmatrix}
-1 & 1\\
-1 & -1
\end{pmatrix}
\begin{pmatrix}
r_R\\
r_L
\end{pmatrix}
\qquad
\mathrm{mod}\ k.
\end{equation}
The determinant of this map is $2$. Therefore, for odd $k$, the map is invertible modulo $k$, and the two Eguchi--Hanson background labels give a complete parametrization of the closed $\mathbb Z_k$ backgrounds on $S^2\times S^2$. 
  \bibliography{RefEHSD.bib}

\providecommand{\href}[2]{#2}\begingroup\raggedright\begin{thebibliography}{10}

\bibitem{Cardy:1981qy}
J.~L. Cardy and E.~Rabinovici, {\it {Phase structure of Zp models in the
  presence of a {\ensuremath{\theta}} parameter}},  {\em Nucl. Phys. B} {\bf
  205} (1982) 1--16.

\bibitem{Cardy:1981fd}
J.~L. Cardy, {\it {Duality and the {\ensuremath{\theta}} parameter in Abelian
  lattice models}},  {\em Nucl. Phys. B} {\bf 205} (1982) 17--26.

\bibitem{Shapere:1988zv}
A.~D. Shapere and F.~Wilczek, {\it {Selfdual Models with Theta Terms}},  {\em
  Nucl. Phys. B} {\bf 320} (1989) 669--695.

\bibitem{Anosova:2022yqx}
M.~Anosova, C.~Gattringer, N.~Iqbal, and T.~Sulejmanpasic, {\it {Phase
  structure of self-dual lattice gauge theories in 4d}},  {\em JHEP} {\bf 06}
  (2022) 149, [\href{http://arxiv.org/abs/2203.14774}{{\tt arXiv:2203.14774}}].

\bibitem{Witten:1995gf}
E.~Witten, {\it {On $S$-duality in Abelian gauge theory}},  {\em Selecta Math.}
  {\bf 1} (1995) 383, [\href{http://arxiv.org/abs/hep-th/9505186}{{\tt
  hep-th/9505186}}].

\bibitem{Verlinde:1995mz}
E.~P. Verlinde, {\it {Global aspects of electric - magnetic duality}},  {\em
  Nucl. Phys. B} {\bf 455} (1995) 211--228,
  [\href{http://arxiv.org/abs/hep-th/9506011}{{\tt hep-th/9506011}}].

\bibitem{Metlitski:2015yqa}
M.~A. Metlitski, {\it {$S$-duality of $u(1)$ gauge theory with $\theta =\pi$ on
  non-orientable manifolds: Applications to topological insulators and
  superconductors}},  \href{http://arxiv.org/abs/1510.05663}{{\tt
  arXiv:1510.05663}}.

\bibitem{Vafa:1994tf}
C.~Vafa and E.~Witten, {\it {A Strong coupling test of S duality}},  {\em Nucl.
  Phys. B} {\bf 431} (1994) 3--77,
  [\href{http://arxiv.org/abs/hep-th/9408074}{{\tt hep-th/9408074}}].

\bibitem{Aharony:2013hda}
O.~Aharony, N.~Seiberg, and Y.~Tachikawa, {\it {Reading between the lines of
  four-dimensional gauge theories}},  {\em JHEP} {\bf 08} (2013) 115,
  [\href{http://arxiv.org/abs/1305.0318}{{\tt arXiv:1305.0318}}].

\bibitem{Choi:2022rfe}
Y.~Choi, H.~T. Lam, and S.-H. Shao, {\it {Noninvertible Time-Reversal
  Symmetry}},  {\em Phys. Rev. Lett.} {\bf 130} (2023), no.~13 131602,
  [\href{http://arxiv.org/abs/2208.04331}{{\tt arXiv:2208.04331}}].

\bibitem{Hayashi:2022fkw}
Y.~Hayashi and Y.~Tanizaki, {\it {Non-invertible self-duality defects of
  Cardy-Rabinovici model and mixed gravitational anomaly}},  {\em JHEP} {\bf
  08} (2022) 036, [\href{http://arxiv.org/abs/2204.07440}{{\tt
  arXiv:2204.07440}}].

\bibitem{Kaidi:2022uux}
J.~Kaidi, G.~Zafrir, and Y.~Zheng, {\it {Non-invertible symmetries of $
  \mathcal{N} $ = 4 SYM and twisted compactification}},  {\em JHEP} {\bf 08}
  (2022) 053, [\href{http://arxiv.org/abs/2205.01104}{{\tt arXiv:2205.01104}}].

\bibitem{Hofman:2024oze}
D.~M. Hofman and S.~Vitouladitis, {\it {Generalised symmetries and
  state-operator correspondence for nonlocal operators}},  {\em JHEP} {\bf 02}
  (2025) 061, [\href{http://arxiv.org/abs/2406.02662}{{\tt arXiv:2406.02662}}].

\bibitem{Shao:2025qvf}
S.-H. Shao and S.~Zhong, {\it {Where non-invertible symmetries end: twist
  defects for electromagnetic duality}},  {\em JHEP} {\bf 01} (2026) 118,
  [\href{http://arxiv.org/abs/2509.21279}{{\tt arXiv:2509.21279}}].

\bibitem{Atiyah:1989vu}
M.~Atiyah, {\it {Topological quantum field theories}},  {\em Inst. Hautes
  Etudes Sci. Publ. Math.} {\bf 68} (1989) 175--186.

\bibitem{Eguchi:1980jx}
T.~Eguchi, P.~B. Gilkey, and A.~J. Hanson, {\it {Gravitation, Gauge Theories
  and Differential Geometry}},  {\em Phys. Rept.} {\bf 66} (1980) 213.

\bibitem{Kronheimer:1989zs}
P.~B. Kronheimer, {\it {The construction of ALE spaces as
  hyper-K\"ahlerquotients}},  {\em J. Diff. Geom.} {\bf 29} (1989), no.~3
  665--683.

\bibitem{PeterBKronheimer:1990zmj}
P.~B. Kronheimer and H.~Nakajima, {\it {Yang-Mills instantons on ALE
  gravitational instantons}},  {\em Math. Ann.} {\bf 288} (1990), no.~1
  263--307.

\bibitem{Katz:1997eq}
S.~Katz, P.~Mayr, and C.~Vafa, {\it {Mirror symmetry and exact solution of 4-D
  N=2 gauge theories: 1.}},  {\em Adv. Theor. Math. Phys.} {\bf 1} (1998)
  53--114, [\href{http://arxiv.org/abs/hep-th/9706110}{{\tt hep-th/9706110}}].

\bibitem{Johnson:1996py}
C.~V. Johnson and R.~C. Myers, {\it {Aspects of type IIB theory on ALE
  spaces}},  {\em Phys. Rev. D} {\bf 55} (1997) 6382--6393,
  [\href{http://arxiv.org/abs/hep-th/9610140}{{\tt hep-th/9610140}}].

\bibitem{Nakajima:1994nid}
H.~Nakajima, {\it {Instantons on ALE spaces, quiver varieties, and Kac-Moody
  algebras}},  {\em Duke Math. J.} {\bf 76} (1994), no.~2 365--416.

\bibitem{Anber:2025gvb}
M.~M. Anber, {\it {Gauging the Standard Model 1-form symmetry via gravitational
  instantons}},  {\em JHEP} {\bf 02} (2026) 225,
  [\href{http://arxiv.org/abs/2509.22788}{{\tt arXiv:2509.22788}}].

\bibitem{Anber:2025dxy}
M.~M. Anber, {\it {Anomalies on ALE spaces and phases of gauge theory}},
  \href{http://arxiv.org/abs/2512.11970}{{\tt arXiv:2512.11970}}.

\bibitem{Bianchi:1996zj}
M.~Bianchi, F.~Fucito, G.~Rossi, and M.~Martellini, {\it {Explicit construction
  of Yang-Mills instantons on ALE spaces}},  {\em Nucl. Phys. B} {\bf 473}
  (1996) 367--404, [\href{http://arxiv.org/abs/hep-th/9601162}{{\tt
  hep-th/9601162}}].

\bibitem{Gaiotto:2014kfa}
D.~Gaiotto, A.~Kapustin, N.~Seiberg, and B.~Willett, {\it {Generalized Global
  Symmetries}},  {\em JHEP} {\bf 02} (2015) 172,
  [\href{http://arxiv.org/abs/1412.5148}{{\tt arXiv:1412.5148}}].

\bibitem{Eguchi:1978gw}
T.~Eguchi and A.~J. Hanson, {\it {Selfdual Solutions to Euclidean Gravity}},
  {\em Annals Phys.} {\bf 120} (1979) 82.

\bibitem{DiFrancesco:1997nk}
P.~Di~Francesco, P.~Mathieu, and D.~Senechal, {\em {Conformal Field Theory}}.
\newblock Graduate Texts in Contemporary Physics. Springer-Verlag, New York,
  1997.

\bibitem{Gibbons:1978tef}
G.~W. Gibbons and S.~W. Hawking, {\it {Gravitational Multi - Instantons}},
  {\em Phys. Lett. B} {\bf 78} (1978) 430.

\bibitem{Donaldson1986Connections}
S.~K. Donaldson, {\it Connections, cohomology and the intersection forms of
  4-manifolds},  {\em Journal of Differential Geometry} {\bf 24} (1986), no.~3
  275--341.

\bibitem{Anselmi:1993sm}
D.~Anselmi, M.~Billo, P.~Fre, L.~Girardello, and A.~Zaffaroni, {\it {ALE
  manifolds and conformal field theories}},  {\em Int. J. Mod. Phys. A} {\bf 9}
  (1994) 3007--3058, [\href{http://arxiv.org/abs/hep-th/9304135}{{\tt
  hep-th/9304135}}].

\bibitem{Witten:2025ayw}
E.~Witten, {\it {Bras and kets in Euclidean path integrals}},  {\em Beijing J.
  Pure Appl. Math.} {\bf 3} (2026), no.~1 1--34,
  [\href{http://arxiv.org/abs/2503.12771}{{\tt arXiv:2503.12771}}].

\bibitem{MilnorHusemoller1973}
J.~W. Milnor and D.~Husemoller, {\em Symmetric Bilinear Forms}, vol.~73 of {\em
  Ergebnisse der Mathematik und ihrer Grenzgebiete}.
\newblock Springer-Verlag, New York, 1973.

\bibitem{Seiberg:2018ntt}
N.~Seiberg, Y.~Tachikawa, and K.~Yonekura, {\it {Anomalies of Duality Groups
  and Extended Conformal Manifolds}},  {\em PTEP} {\bf 2018} (2018), no.~7
  073B04, [\href{http://arxiv.org/abs/1803.07366}{{\tt arXiv:1803.07366}}].

\bibitem{Anderson:1990sgi}
M.~T. Anderson, {\it Short geodesics and gravitational instantons},  {\em
  Journal of Differential Geometry} {\bf 31} (1990), no.~1 265--275.

\bibitem{Kapustin:2013qsa}
A.~Kapustin and R.~Thorngren, {\it {Topological Field Theory on a Lattice,
  Discrete Theta-Angles and Confinement}},  {\em Adv. Theor. Math. Phys.} {\bf
  18} (2014), no.~5 1233--1247, [\href{http://arxiv.org/abs/1308.2926}{{\tt
  arXiv:1308.2926}}].

\bibitem{Witten:2003ya}
E.~Witten, {\it {SL(2,Z) action on three-dimensional conformal field theories
  with Abelian symmetry}},  in {\em {From Fields to Strings: Circumnavigating
  Theoretical Physics: A Conference in Tribute to Ian Kogan}}, pp.~1173--1200,
  7, 2003.
\newblock \href{http://arxiv.org/abs/hep-th/0307041}{{\tt hep-th/0307041}}.

\bibitem{Zucchini:2003in}
R.~Zucchini, {\it {Four-dimensional Abelian duality and SL(2,Z) action in
  three-dimensional conformal field theory}},  {\em Adv. Theor. Math. Phys.}
  {\bf 8} (2004), no.~5 895--936,
  [\href{http://arxiv.org/abs/hep-th/0311143}{{\tt hep-th/0311143}}].

\bibitem{Kapustin:2009av}
A.~Kapustin and M.~Tikhonov, {\it {Abelian duality, walls and boundary
  conditions in diverse dimensions}},  {\em JHEP} {\bf 11} (2009) 006,
  [\href{http://arxiv.org/abs/0904.0840}{{\tt arXiv:0904.0840}}].

\bibitem{Seiberg:2016gmd}
N.~Seiberg, T.~Senthil, C.~Wang, and E.~Witten, {\it {A Duality Web in 2+1
  Dimensions and Condensed Matter Physics}},  {\em Annals Phys.} {\bf 374}
  (2016) 395--433, [\href{http://arxiv.org/abs/1606.01989}{{\tt
  arXiv:1606.01989}}].

\bibitem{hatcher2002algebraic}
A.~Hatcher, {\em Algebraic Topology}.
\newblock Cambridge University Press, Cambridge, 2002.

\end{thebibliography}\endgroup
  
  \bibliographystyle{JHEP}
  \end{document}